%% file: hvphighQ2.tex
\documentclass[12pt,tightenlines,preprintnumbers,superscriptaddress,floatfix,nofootinbib]{revtex4}
\usepackage[T1]{fontenc}
\usepackage[utf8]{inputenc}
\usepackage{lmodern}
\usepackage[english]{babel}
\usepackage{graphicx}
\usepackage{amsmath,mathtools,amssymb}
\usepackage{hyperref}
\usepackage{tabularx}
\usepackage{hhline}
\usepackage{booktabs}
\usepackage[dvipsnames]{xcolor}
\usepackage{tikz}
\usepackage{bm}
\usetikzlibrary{decorations.pathreplacing,decorations.pathmorphing}

\usepackage{siunitx}
\usepackage{soul}

\input{definitions}

\newcommand{\dint}{\mathrm d}
\newcommand{\sgn}{\mathrm {sgn}}
\newcommand{\g}{\gamma}
\usepackage{physics}
\usepackage{siunitx}

\begin{document}

\title{Vacuum correlators at short distances from lattice QCD}

\preprint{MITP/21-032, CERN-TH-2021-100}
      
\author{Marco C{\`e}} 
\affiliation{Theoretical Physics Department, 
CERN, 
CH-1211 Geneva 23, Switzerland}

\author{Tim Harris} 
\affiliation{School of Physics and Astronomy,\\
University of Edinburgh, EH9 3JZ, UK}

\author{Harvey B.\ Meyer} 
\affiliation{Helmholtz~Institut~Mainz,
Johannes Gutenberg-Universit\"at Mainz,
D-55099 Mainz, Germany}
\affiliation{GSI Helmholtzzentrum f\"ur Schwerionenforschung, Darmstadt, Germany}
\affiliation{PRISMA$^+$ Cluster of Excellence \& Institut f\"ur Kernphysik,
Johannes Gutenberg-Universit\"at Mainz,
D-55099 Mainz, Germany}

\author{Arianna Toniato} 
\affiliation{PRISMA$^+$ Cluster of Excellence \& Institut f\"ur Kernphysik,
Johannes Gutenberg-Universit\"at Mainz,
D-55099 Mainz, Germany}

\author{Csaba T\"or\"ok} 
\affiliation{PRISMA$^+$ Cluster of Excellence \& Institut f\"ur Kernphysik,
Johannes Gutenberg-Universit\"at Mainz,
D-55099 Mainz, Germany}

\begin{abstract}
  Non-perturbatively computing the hadronic vacuum polarization at large photon virtualities
  and making contact with perturbation theory enables a precision determination of the electromagnetic coupling
  at the $Z$ pole, which enters global electroweak fits. In order to achieve this goal ab initio using lattice QCD, one
  faces the challenge that, at the short distances which dominate the observable, discretization errors are hard to control.
  Here we address challenges of this type with the help of static screening correlators in the high-temperature phase of QCD,
  yet without incurring any bias. The idea is motivated by the observations that
  (a) the cost of high-temperature simulations is typically much lower than their vacuum counterpart,
  and (b) at distances $x_3$ far below the inverse temperature $1/T$,
  the operator-product expansion guarantees  the thermal correlator of two local currents
  to deviate from the vacuum correlator by a relative amount that is power-suppressed in $(x_3\:T)$.
  The method is first investigated in lattice perturbation theory, where we point out the appearance of an O$(a^2 \log(1/a))$
  lattice artifact in the vacuum polarization with a prefactor that we calculate.
  It is then applied to non-perturbative lattice QCD data with two dynamical flavors of quarks.
  Our lattice spacings range down to 0.049\,fm for the vacuum simulations and down to 0.033\,fm for the simulations performed
  at a temperature of 250\,MeV.
\end{abstract}

\maketitle

\section{Introduction}

Many phenomenologically interesting observables are defined in terms
of QCD vacuum correlators involving two or more local fields
integrated over their Euclidean positions.  For example,
the hadronic vacuum polarization function $\widehat\Pi(Q^2)$, which determines
the leading hadronic contribution to the running of the
electromagnetic coupling and the muon anomalous magnetic moment $(g-2)_\mu$, the
$\pi^0\to\gamma^*\gamma^*$ transition form factor of the pion and the
hadronic light-by-light contribution to $(g-2)_\mu$
are expressed in such a way. Thus in many cases, vacuum correlators 
represent crucial input for precision tests of the Standard Model.
Lattice QCD provides ab initio determinations of these vacuum
correlators; see e.g.\ Refs.~\cite{Burger:2015lqa,Francis:2015grz,Gerardin:2019vio,Aoyama:2020ynm}
for the applications above.

However, these integrated quantities contain contributions
corresponding to the fields being close together, a regime which can
lead to large cutoff effects.  The standard tool to investigate the
asymptotic approach to the continuum limit of correlation functions is
Symanzik's effective field
theory~\cite{Symanzik:1983dc,Luscher:1998pe,Weisz:2010nr}. A
complication for the aforementioned observables is that the on-shell
improvement programme is not sufficient to guarantee rapid convergence
toward the continuum limit.  In this work, we   
propose to compute the short-distance contribution to vacuum
correlators by making use of static screening correlators from QCD at finite
temperature, which can significantly reduce the cost of obtaining a
robust continuum limit.  As the short-distance contribution has little
sensitivity to the temperature, the bulk of this contribution can be
computed using particularly small lattice spacings in the 
high-temperature phase of QCD, where the cost of the simulations is much reduced, and
only a small remainder needs to be computed using the vacuum
ensembles. Just as importantly, the cutoff effects on the remainder can be arranged to be 
parametrically smaller than those of the observable computed on thermal gauge ensembles.
Furthermore, we note that in certain cases, there is a
logarithmic enhancement of the cutoff effects on the short-distance
contribution at leading order in perturbation theory, in contrast to
the modification of cutoff effects by logarithms affecting on-shell correlators,
which only appears beyond the free-field theory level~\cite{Husung:2019ytz}.  
The longer-distance contribution involves the currents at
physical separations, at which on-shell improvement can safely be applied directly to the vacuum correlators.

The fact that the leading thermal effect on the correlator at short distances $x_3$ is suppressed by several powers
of $(x_3\:T)$ allows for a strategy to compute correlators at extremely high momentum scales $|Q|$.
We make a fairly concrete proposal in this direction in Section~\ref{sec:concl} for how to compute the hadronic contribution
to the running of the electroweak coupling constants up to the $Z$-boson mass.
The basic idea is to compute this contribution by increasing the momentum scale by factors of two,
always using thermal QCD ensembles with $|Q|/T$ sufficiently large that the thermal effects are small corrections
computable to a systematically improvable accuracy. The idea thus has common aspects with the `step-scaling' idea
introduced in the lattice field theory context in~\cite{Luscher:1991wu}, and also with the application in heavy-quark physics presented in Ref.~\cite{deDivitiis:2003iy}.

In the following section, we outline the strategy for computing short-distance observables
using auxiliary finite-temperature ensembles,
and provide parametric estimates for the optimal choice of lattice parameters.
We examine in Section~\ref{sec:pert} the case of the vector current correlator in
the free theory, which suggests that the thermal effects are
guaranteed to be small at sufficiently small separations of the
currents.  This is confirmed by the operator product expansion, carried out at
leading and next-to-leading order in Appendix~\ref{app:OPE}.
In Section~\ref{sec:nf2}, we test the
strategy on vector current correlators with $N_\mathrm{f}=2$ Wilson
fermions with thermal ensembles with a temperature $T=250$ MeV, where
we reach lattice spacings of $a\approx0.033$fm.  This provides a more
controlled continuum limit of the short-distance contribution to the
vacuum polarization at a much reduced cost.  Finally,
Section~\ref{sec:concl} summarizes our findings and describes
the idea to compute the running of electroweak couplings
to very high energy using sequences of ensembles of growing temperature.
The concrete setup  suggested  is tested in the free-theory context in Appendix~\ref{sec:Delta2Pi}.
The other appendices contain technical details of the analytic calculations.

\section{Definitions \& general idea \la{sec:defs}}
To be specific, in this study we concentrate on two observables 
which are defined in terms of the integral over
a Euclidean correlator weighted by a known kernel and are closely related to the hadronic
vacuum polarization function $\Pi(Q^2)$.
The Adler function~\cite{Bernecker:2011gh}
\begin{align}
    \label{eq:adler}
    D(Q^2) &= 12\pi^2 Q^2\frac{\mathrm{d}\Pi}{\mathrm{d}Q^2} = \int_0^\infty\mathrm{d}x_0\; K(x_0,Q^2)G(x_0),\\
    \label{eq:kernel}
    K(x_0,Q^2) &= \frac{12\pi^2}{Q^2}\Big[2- 2\cos(Qx_0) - Qx_0\sin(Qx_0)\Big],
\end{align}
parametrizes the running of the hadronic contribution to the electromagnetic
coupling at spacelike $q^2=-Q^2<0$, and its derivative at $Q^2=0$
\begin{align}
    \frac{D'(0)}{\pi^2} &= \int_0^\infty\mathrm{d}x_0\,x_0^4\,G(x_0),
    \label{eq:x04moment}
\end{align}
determines the anomalous magnetic moment of a lepton in the limit of vanishing
lepton mass, $m_l$.
Both of these quantities receive contributions from all non-zero
time-separations of the current correlator
\begin{align}
    G(x_0) &= -\int \mathrm{d}^3x\; \langle
        J^\mathrm{em}_1(x) J^\mathrm{em}_1(0)
        \rangle,
    \label{eq:G}
\end{align}
where the (continuum) electromagnetic current is defined as
\begin{align}
    J^\mathrm{em}_\mu(x) &= \sum_f Q_f \bar\psi^f(x)\gamma_\mu\psi^f(x),
    \label{eq:emcorr}
\end{align}
and $Q_f$ is the electric charge of quark flavour
$f=\mathrm{u},\mathrm{d},\mathrm{s},\ldots$ and the matrices $\gamma_\mu$
satisfy the Euclidean Dirac algebra
$\{\gamma_\mu,\gamma_\nu\}=2\delta_{\mu\nu}$.
The kernels for both the Adler function and the anomalous lepton magnetic moment $(g-2)_\mu$ in
the time-momentum representation~\cite{Bernecker:2011gh} coincide with the
fourth moment for small enough $x_0$.
In fact, in the case of $(g-2)_\mu$, the
kernel agrees with the fourth moment at the percent level up to distances of about $0.5$\,fm.
Therefore, both the qualitative and quantitative results for the fourth moment
are relevant for the controlled determination of the short-distance contribution to
the hadronic vacuum polarization in the muon anomalous magnetic moment.

One may wonder whether lattice QCD estimates of these physical quantities
suffer from uncontrolled systematic effects arising from small separations
between the currents, even if this contribution itself is suppressed by the
short-distance behaviour of the kernel.
Indeed, for a lattice estimator of the current
correlator which has the expansion in the lattice spacing $a$ given by
\begin{align}
    \mathcal G(x_0,a) &= G(x_0) + a\mathcal G_1(x_0,a) + a^2 \mathcal G_2(x_0,a) + \ldots,
    \label{eq:sym_corr}
\end{align}
power counting suggests that its cutoff effects become parametrically large as
$x_0$ becomes small~\cite{DellaMorte:2008xb}
\begin{align}
    a^n\mathcal G_n(x_0,a) &= \mathrm{const.}\times\left({a}/{x_0}\right)
        ^nG(x_0)+\ldots,
    \label{eq:latart}
\end{align}
up to logarithmic corrections~\cite{Husung:2019ytz}, and where we assume the
continuum limit $G(x_0)$ exists after proper renormalization, if required.
Even if the bulk of the lattice artifacts does not necessarily arise from the short-distance
contribution, the breakdown of the Symanzik expansion inevitably leads to
scaling violations in the continuum limit which can be of practical concern,
especially given the subpercent precision aimed at in the context of $(g-2)_\mu$.

This situation is similar to the typical window-problem encountered in lattice
QCD where the appearance of an external scale, such as $Q^2$, needs to be
accommodated within the ultra-violet and infra-red cutoffs imposed by the
lattice spacing $a$ and lattice size $L$,
\begin{align}
    a\ll Q^{-1}\ll L.
    \label{eq:window}
\end{align}
In renormalization problems, one has the freedom to remove one of these
restrictions by linking the external scale to the physical volume, which
eliminates one constraint of the window and allows simulations to proceed with
tractable problem sizes.
For hadronic observables, where the physical volume must remain large, we may
however choose to compute an observable, or part of it, in a simulation with
different physical parameters provided that we properly account for the 
correction.

\begin{figure}[tp]
    \centering
    \includegraphics[scale=0.6]{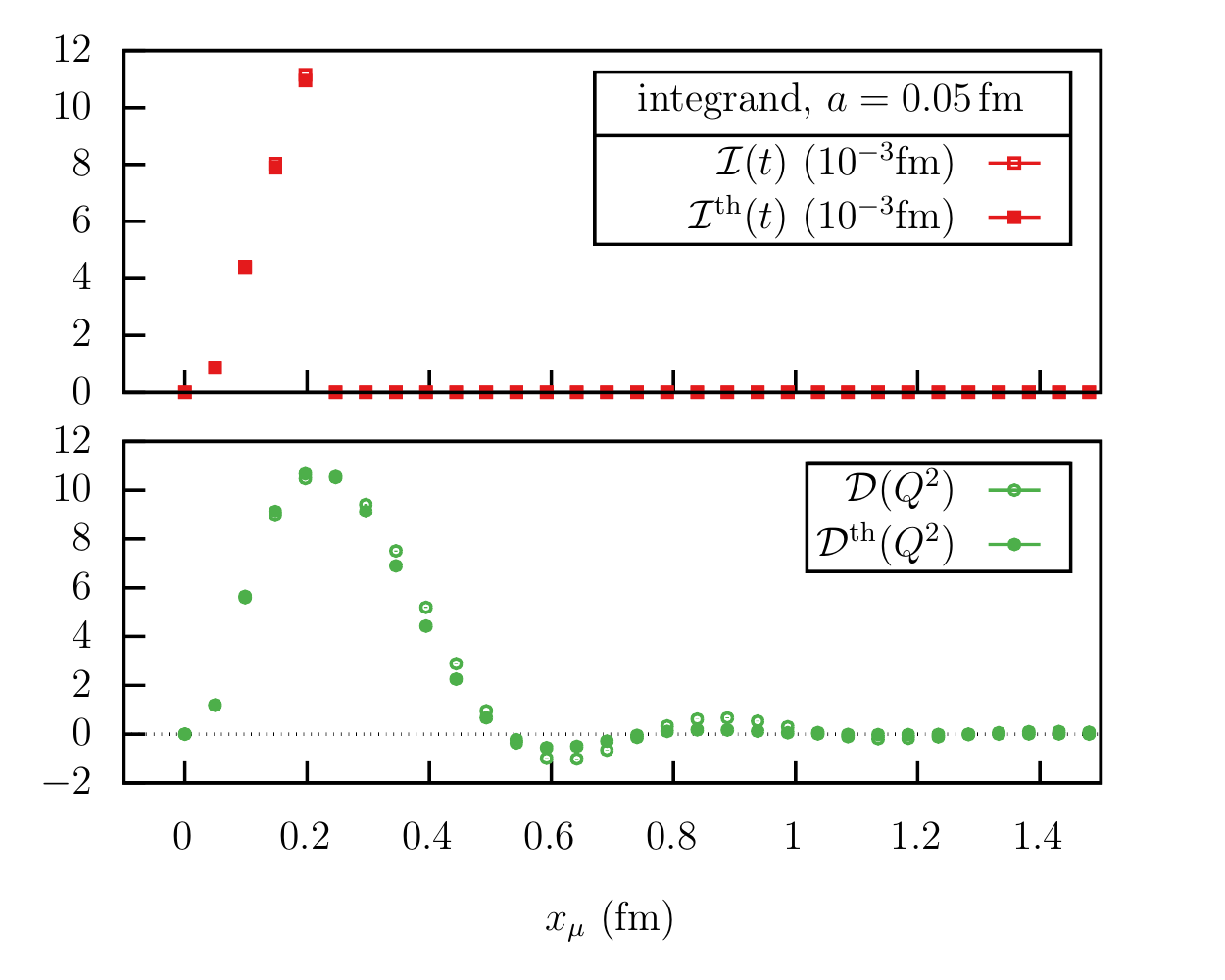}
    \includegraphics[scale=0.6]{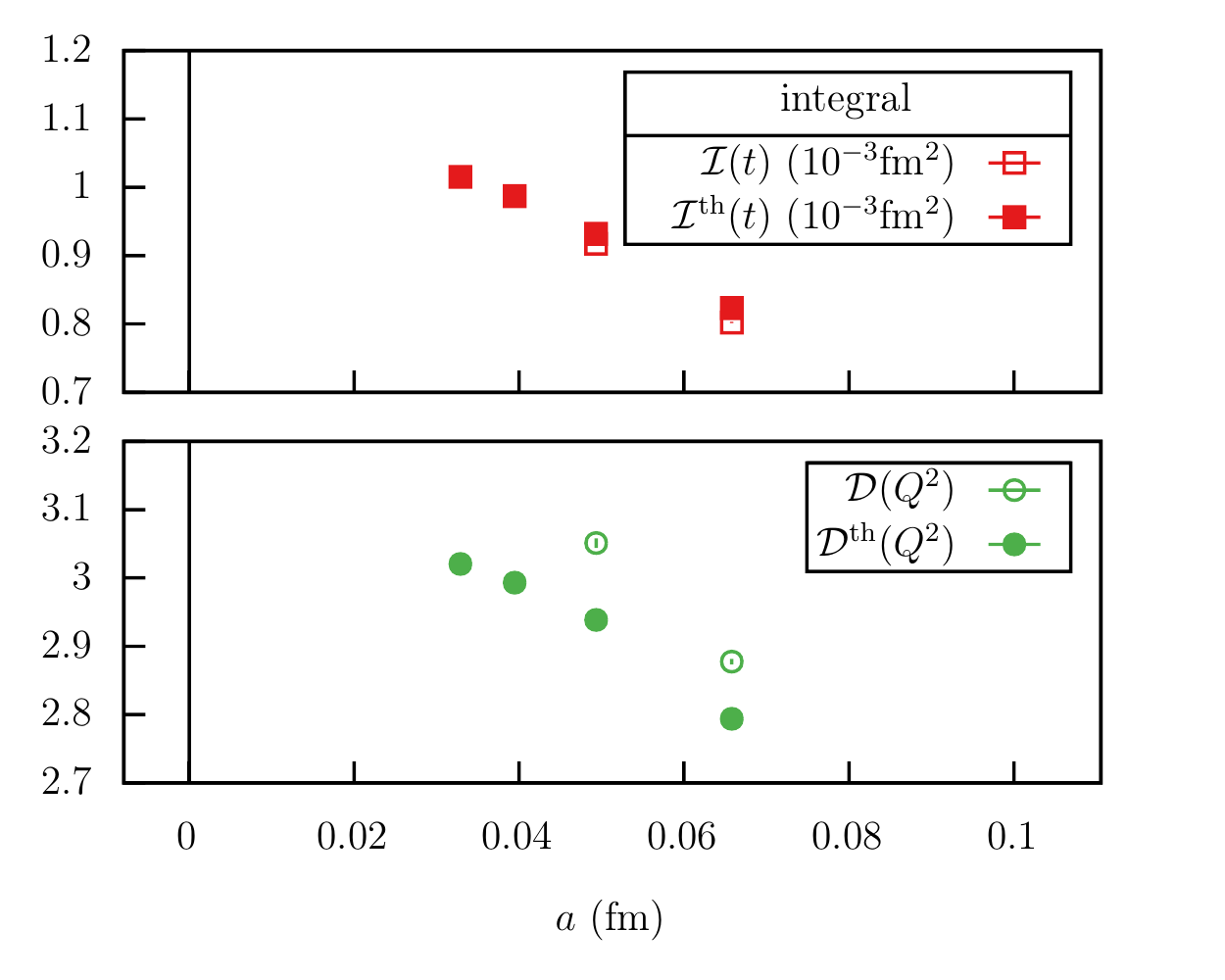}
    \caption{
       Left panel: Integrand of the fourth moment of the correlator (red squares)
       and the Adler function (green circles) at fixed lattice spacing.
       The vacuum and thermal observables are shown with open and filled
       symbols, respectively.
       Right panel: Corresponding integral as a function of the lattice spacing is shown,
       which illustrates that the thermal contribution
       with $T=250$ MeV accounts for most of the signal.
    }
    \label{fig:nf2_compare}
\end{figure}

In particular, our strategy proposes to use the static screening correlator at finite temperature $T$,
\begin{align}
    G^\mathrm{th}(x_3) &= -\int \mathrm{d}x_0\,\mathrm{d}x_1\,\mathrm{d}x_2\; \langle
        J^\mathrm{em}_1(x) J^\mathrm{em}_1(0)
        \rangle_{T},
    \label{eq:emscreening}
\end{align}
which is a function of the spatial separation $x_3$ of the currents and depends
on the temperature $T$. We choose the latter to be on the order of the QCD scale, in the chirally-restored phase.
We define the contribution up to $t$ to the integral appearing in
Eq.~\eqref{eq:x04moment} for the vacuum and thermal correlators,
\begin{align}
    I(\xmax) &= \int_0^\xmax \mathrm dx_0\, x_0^4\,G(x_0),\qquad
    I^\mathrm{th}(\xmax) = \int_0^\xmax \mathrm dx_3\, x_3^4\,G^\mathrm{th}(x_3),
    \label{eq:integral}
\end{align}
together with lattice estimators $\mathcal I(t,a)$ and its thermal counterpart
$\mathcal I^\mathrm{th}(t,a)$ defined precisely in the following subsection.

Our strategy is based on the idea that the quantities $I(\xmax)$ and
$I^\mathrm{th}(\xmax)$ are in some sense very similar. The
operator-product expansion (OPE), which is presented in detail in 
Appendix~\ref{app:OPE}, can be invoked to make this statement precise for $\xmax
T\ll 1$: the difference of the two quantities is suppressed by $(\xmax
T)^3$ relative to the quantities themselves.  It is instructive to
compare the thermal and the vacuum correlators in a representative
lattice QCD calculation.  The left panel of
Figure~\ref{fig:nf2_compare} depicts the integrand of
Eq.~\eqref{eq:integral} with $t= 0.2$ fm for the vacuum (open) and
thermal (filled) squares at fixed lattice spacing $a\approx 0.05$ fm,
which illustrates that the thermal effects are indeed suppressed for
these distances in the $\Nf=2$ theory when $T=250$ MeV, corresponding
to $\xmax T= 0.25$.  The analogous integrands for the Adler function at
the large virtuality of $Q=2.36$\;GeV are shown as well, which illustrate how it
is also dominated by the correlation function at short distances.
The right-hand panel shows the corresponding integrals as a function
of the lattice spacing, which illustrates that more than 95\% of the
signal is accounted for by the thermal observable.  The benefit of
using $I^\mathrm{th}(\xmax)$ as a proxy for $I(\xmax)$ is that, in the case
illustrated in Figure~\ref{fig:nf2_compare},
the finite-temperature ensemble has a factor eight fewer lattice sites
than its vacuum counterpart due to its shorter time extent, which naively allows a span of a
factor of $8^{1/4}\approx 1.68$ in the lattice spacing to be achieved
for the thermal observable before the cost of obtaining the latter
becomes comparable to the vacuum calculation.

Thus for a suitable choice of $\xmax\ll 1/T$, we expect the bulk of the
short-distance contribution to be given by the thermal component
$I^\mathrm{th}(\xmax)$, whose continuum limit can be obtained accurately thanks
to the smaller lattice spacings accessible at finite temperature.
This suggests an improved estimator for the vacuum observable 
\begin{align}
    \widehat{\mathcal I}(\xmax,a) &=
    I^\mathrm{th}(\xmax) + [\mathcal I(\xmax,a) - \mathcal I^\mathrm{th}(\xmax,a)],
    \label{eq:imp}
\end{align}
where the first term on the right-hand side is the continuum estimate of the thermal observable,
obtained using particularly fine lattice spacings available at high
temperature. 
  The remainder in brackets is small and of the form ${\rm
  const.}\times t^5(1+{\rm O}(t))$, where the constant is dominated by
a momentum scale on the order of temperature.  It is worth recording
the parametric size of the cutoff effects on both terms: for the first
term, the cutoff effects are of order $a^2$, while for the remainder,
they are of order $ a^2 (Tt)^3$ in the O($a$)-improved
theory\footnote{In the unimproved theory, there are also cutoff
effects of order $aT^4t^5$, but none of order $aT^3t^4$. The former are
still small compared to the cutoff effects on the first term, provided
$(Tt)^5 \ll aT$, which is certainly the case in our numerical
application of Section~\ref{sec:nf2}.}.  Thus, as long as the ratio of the
lattice spacings used in the thermal theory to those used in the vacuum does not
become as small as $(tT)^{3/2}$, which is the regime we have in mind, the
cutoff effects on $\mathcal I^\mathrm{th}(\xmax,a)$ are parametrically larger than the
effects on the remainder. The upshot is that if $I(t)$ is obtained as the continuum limit of
$\widehat{\mathcal I}(\xmax,a)$, the dominant part of the systematic error associated with cutoff effects 
comes from obtaining $I^\mathrm{th}(\xmax)$, where one profits from being able to reach lattice spacings below 0.05\,fm
in the chirally-restored phase at a moderate computational cost.

A further aspect which is specific to Wilson fermions is that the
on-shell improvement of the vector current via the derivative of the
tensor current~\cite{Sint:1997jx} only contributes a term of order
$am_{\rm q}$ in the chiral restored phase of QCD; such terms are of a
size comparable to the O($a^2$) terms for the lattice spacings
employed in Section~\ref{sec:nf2}.  This represents a further
advantage of obtaining the bulk of $\widehat{\mathcal I}(\xmax,a)$
from the chiral restored phase.

Since the main focus is then on obtaining $I^\mathrm{th}(\xmax)$, it
is worth studying the approach to the continuum for this
short-distance quantity in lattice perturbation theory.  The
leading-order calculation is presented in Section~\ref{sec:pert}.  It
turns out that a logarithmic enhancement of the O($a^2$) cutoff
effects arises, with a calculable coefficient which applies both to
$I^\mathrm{th}(\xmax)$ and $I(\xmax)$.  This enhancement appears with
a positive (unit) power of the logarithm, unlike the known logarithmic
dependence on the lattice spacing due to the running coupling, which
appears first at one-loop level~\cite{Husung:2019ytz}.
By contrast, the $\mathrm{O}(a^2)$ cutoff effect enhanced by the factor $\log(1/a)$
cancels out in the improved observable (\ref{eq:imp}).

In order to fully control the short-distance thermal contribution, or
to reach very high momenta in the hadronic vacuum polarization, it may
be necessary to iterate the procedure of Eq.~\eqref{eq:imp} using a
series of higher temperatures to compute short-distance
contributions. We return to this question in Section~\ref{sec:concl}.
Although other options are certainly available, it is particularly
convenient to use the temperature as a control parameter to compute
the short-distance contribution, since we can use existing knowledge about
high-temperature correlators and apply well-understood theoretical tools like
the operator product expansion.

\subsection{Definitions of lattice observables}
\label{sect:lattice_obs}

In order to set up the notation for the following sections, we define here the
lattice observables for the theory of $\Nf=2$ Wilson fermions.
In this work we investigate the isovector vector current correlator, which consists
of a single Wick contraction. This correlator makes the dominant contribution to the
hadronic vacuum polarization in the muon $g-2$. 
In the vacuum case, we formulate the correlator at vanishing spatial momentum as a function
of Euclidean time,
\begin{align}
    \mathcal G_{\mu\nu}(x_0) &= 
    Z_\mathrm{V}(1+am_\mathrm{q}b_\mathrm{V}) a^3\sum_{\bm x}\langle \widetilde V_\mu(x) V_\nu(0)\rangle\;,
    \label{eq:gvac}
\end{align}
where the bare local vector current is defined as 
\begin{align}
    V_\mu(x) &= \bar \Psi(x)\frac{\tau_3}{\sqrt{2}}\gamma_\mu\Psi(x)
    \label{eq:localcurrent}
\end{align}
with $\Psi^\top=(u,d)$, and the exactly-conserved vector current is 
\begin{align}
    \widetilde V_\mu(x) &= \frac{1}{2}\big[
        \bar\Psi(x+a\hat\mu)\frac{\tau_3}{\sqrt{2}}
            (1+\gamma_\mu)U^\dagger_\mu(x)\Psi(x)
        -\bar\Psi(x)\frac{\tau_3}{\sqrt{2}}
            (1-\gamma_\mu)U_\mu(x)\Psi(x+a\hat\mu)\Big].
    \label{eq:conscurrent}
\end{align}
In contrast, the static screening correlator at finite temperature is measured along a spatial direction,
\begin{align}
    \mathcal G^\mathrm{th}_{\mu\nu}(x_3) &=
    Z_\mathrm{V} (1+am_\mathrm{q}b_\mathrm{V})a^3\sum_{x_0}\sum_{x_1,x_2}
    \langle \widetilde V_\mu(x) V_\nu(0)\rangle_T
    \label{eq:gth}
\end{align}
We investigate two observables which are related to the Adler function
Eq.~\eqref{eq:adler}, and the short-distance part of the fourth moment of the correlator
Eq.~\eqref{eq:integral}, which is proportional to the derivative of the vacuum polarization
at zero virtuality.
The short-distance contribution up to $t$ of the fourth moment is
\begin{align}
    \mathcal I(\xmax) &= a\sum_{x_0=a}^{\xmax-a}x_0^4\,\mathcal G(x_0)
    + \frac{a}{2}\xmax^4\,\mathcal G(\xmax), \qquad   \mathcal G\equiv  -\mathcal G_{11}.
    \label{eq:x04moment_lat}
\end{align}
We have used an integration rule consistent with the improvement of the theory,
\emph{e.g.}\ the trapezoidal rule.
For large $Q^2$, the Adler function is dominated by the short-distance
contribution to the integral, and we define a lattice observable which is the
integral up to half the spatial extent $L/2$,
\begin{align}
    \mathcal D(Q^2) &= a\sum_{x_0=a}^{L/2-a}K(x_0,Q^2)\mathcal G(x_0)
    + \frac{a}{2}K(L/2,Q^2)\mathcal G(L/2),
    \label{eq:adler_lat}
\end{align}
where again we have implemented the trapezoidal rule, and $K(x_0,Q^2)$ is
defined in Eq.~\eqref{eq:kernel}.
The thermal observables are defined analogously using the static screening
correlator of Eq.~\eqref{eq:gth}.
The improved estimators are then defined via Eq.~\eqref{eq:imp} where the first
term on the right-hand side is obtained by taking the continuum limit using
thermal ensembles with the available finer lattice spacings.

\subsection{The Symanzik expansion and enhanced lattice artifacts}
\label{sub:symanzik}
As discussed at the beginning of the section (see Eqs.\ (\ref{eq:sym_corr}-\ref{eq:latart})), severe lattice artifacts appear
in the correlation function at short distances.
Here, we demonstrate that integrating over the correlation function with a kernel suppressing the short
distances sufficiently so as to yield a finite continuum limit can result in a parametric enhancement
of lattice artifacts even at leading order in the perturbative expansion.

The Symanzik continuum effective theory can be used to represent the
correlation function on the lattice for $x_0>0$ by considering all irrelevant
counterterms of the action and local operators with the correct dimension and
consistent with the symmetries of the lattice theory.
For example, the lattice artifacts of Eq.~\eqref{eq:latart} can be expressed as
a sum over the matrix elements containing the
counterterms~\cite{Husung:2019ytz}
\begin{align}
    \mathcal G_n(x_0,a) &= \sum_i \bar c^i\{2b_0\bar g^2(\tfrac{1}{a})\}^{\hat\gamma^i}
        \mathcal C^i_n(x_0),
        \qquad\hat\gamma^i=\gamma_0^i/b_0,
    \label{eq:logs}
\end{align}
with coefficients which depend on the (scheme-independent) one-loop anomalous
dimension of the counterterm $\gamma_0^i$ ($b_0$ is the universal one-loop
coefficient of the QCD beta function) and $\bar c^i$ is the matching
coefficient between the Symanzik continuum effective theory and the lattice
theory.
The matrix element $\mathcal C_n^i$ is renormalization-group invariant as the
scale-dependence of the counterterm has been factored out, which gives rise to
a logarithmic dependence on the lattice spacing through the running coupling.

In addition, however, the integral of the correlator from short-distances
results in a logarithmically-enhanced lattice artifact.
In Eq.~\eqref{eq:latart}, the matrix element of any one of
the leading $\mathrm O(a^2)$ counterterms must have by power counting
 the short-distance singularity
\begin{align}
    \mathcal C^i_2(x_0) &= d^i (1/x_0)^{5} + \ldots.
    \label{eq:leadinglog}
\end{align}
This is more singular than the leading continuum correlator, and gives rise to
a logarithmic enhancement of the O($a^2$) lattice artifacts in the $x_0^4$ 
moment, in particular when inserted in the summation of
Eq.~\eqref{eq:x04moment_lat}, using the harmonic number formula.
Thus, assuming the correlator is O($a$)-improved, the leading $\mathrm O(a^2)$
lattice artifacts of the integrated quantity are parametrically enhanced due to the
logarithm appearing with the (positive) unit power.
It is also worth noting that all higher terms in the Symanzik expansion of ${\cal G}(x_0,a)$ contribute at O($a^2$)
after summing over short distances. In particular, even if one improved the
on-shell correlator so that it contained no O($a^2$) artifacts,
the quantity ${\cal I}(x_0,a)$ would contain remnant O($a^2$) lattice artifacts.

The full form of the lattice artifacts given by Eq.~\eqref{eq:logs} suggests
that the coefficient of the logarithmic term could be determined by computing
all of the matching, or improvement, coefficients while the most singular
behaviour of the coefficient function is computable in continuum
perturbation theory, when $x_0\Lambda\ll1$.
In the following section, the coefficient of the logarithmically-enhanced term
is computed at leading order in lattice perturbation theory.

\section{Analysis in leading order of lattice perturbation theory\la{sec:pert}}

As a first test of the idea to use thermal gauge ensembles to better control
the short-distance behaviour of QCD correlators, we apply it in the framework of leading-order
lattice perturbation theory, \emph{i.e.}\ in the theory of non-interacting quarks.
In the limit of short distances, the QCD correlation functions are well approximated
by their perturbative values, and we therefore expect the free theory
to provide valuable insight on the general applicability of the method.
Moreover, the continuum values being known in this context,
quantitative statements can be made about the accuracy of the continuum limit
obtained with the improved estimators in Eq.~\eqref{eq:imp} as compared 
to extrapolating directly the vacuum observables.

One delicate point in the study of the integrated observables defined
in Section~\ref{sect:lattice_obs} is the presence, already at the level
of the free theory, of cutoff effects which depend logarithmically on
the lattice spacing.  In the method we propose, the understanding of
these effects is important in view of obtaining an accurate continuum
extrapolation of the thermal observables. On the other hand, the
improved vacuum observables defined as in Eq.~\eqref{eq:imp} are free of the
logarithm predicted by the leading-order calculation, which cancel out in the
subtraction between thermal and vacuum quantities.

\subsection{The vector correlators in the massless theory: lattice formulation}
\label{sect:latt_form}

\begin{figure}[tp]
\center
\includegraphics[width = .65\textwidth]{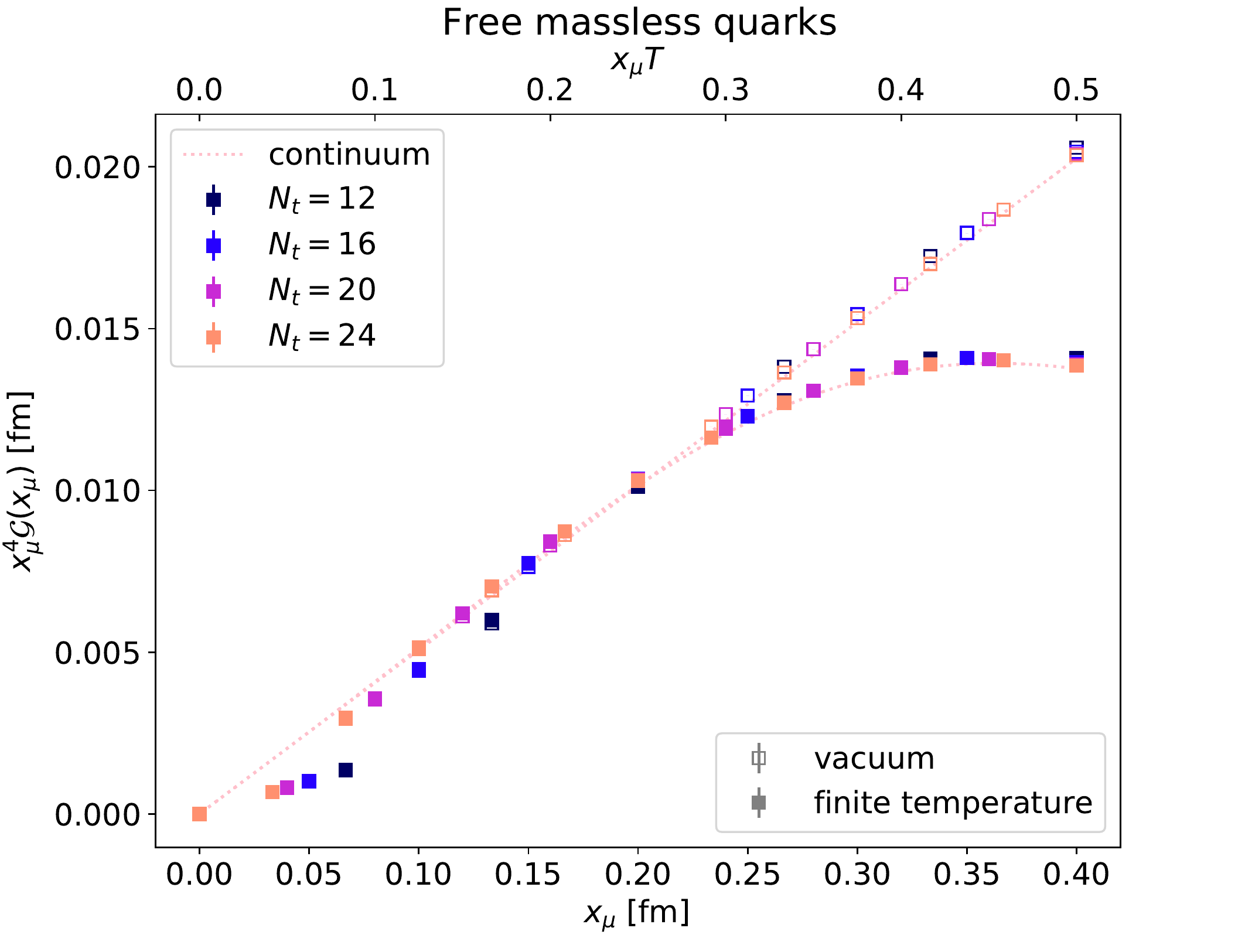}
\caption{Integrand  of the fourth-moment observable
$x_{\mu}^4 \mathcal G (x_{\mu})$ for four thermal lattices 
with $N_t = 12,16,20,24$ (filled symbols) and their vacuum
equivalents (open symbols). The dotted lines represent the 
continuum values of these quantities.
}
\label{fig:free_integrand}
\end{figure}

We consider the theory of non-interacting massless Wilson quarks, defined
on a lattice with infinite spatial volume.
Following Section~\ref{sect:lattice_obs}, we denote the lattice vacuum and
thermal correlation functions by $\mathcal G_{\mu\nu} (x_0)$ and $\mathcal G^{\mathrm{th}}_{\mu\nu} (x_3)$
(Eqs.~(\ref{eq:gvac}) and (\ref{eq:gth}), recalling that $Z_\mathrm{V} = 1$ in the non-interacting case).
Explicit expressions of the free correlators are given in Appendix~\ref{app:free_corr} for the theory with $N_c$ colors.
Here we fix $N_c = 3$, as appropriate for QCD.
We concentrate on the observables $\mathcal I(t)$ and $\mathcal D(Q^2)$, 
as defined in Eqs.~(\ref{eq:x04moment_lat}) and (\ref{eq:adler_lat}).

The analysis is performed at a set of realistic lattice spacings which correspond to those available
in our non-perturbative study of $N_f=2$ QCD (see Table~\ref{tab:nf2}).
In the free massless Lagrangian there are no bare parameters to be
tuned in order to approach the continuum on a line of constant physics, 
given that the mass parameter is only multiplicatively renormalized in
this case.
The fact that the non-interacting massless theory is scale invariant gives us the freedom to assign
to the temperature a value of our choice.
In the thermal case, the lattice spacing and the temperature 
are related by $T = 1/(a N_t)$, where $N_t=L_0/a$ is the number of lattice points 
in the Euclidean-time direction.
As in the case of the interacting ensembles, we consider four
finite-temperature lattices, with $N_t = 12,16,20,24$, and we assign to each of them
the physical temperature $T = 246.25$~MeV, which corresponds to fixing the lattice
extent in the compact direction to $a N_t = 1/T = 0.8$~fm.
This results in the set of lattice spacings $a \approx \{0.07,0.05,0.04,0.03\}$~fm.
Vacuum equivalents of these thermal systems are obtained by assigning
the corresponding physical value of the lattice spacing to a lattice 
 with $N_t = \infty$ (zero temperature).
For simplicity, with an abuse of notation we will sometimes identify the 
vacuum lattices by the value of $N_t$ of their thermal counterpart
(as for example in Figure~\ref{fig:extrap_vac}). 
We analyze $\mathcal I(t)$ for  
$t = 1/(4T) = 0.2$~fm and $t = 1/(2T) = 0.4$~fm, and for the Adler function 
we consider the two virtualities $Q = 3\pi T = 2.32$~GeV and $Q = \pi T/2 = 387$~MeV.
We are mostly interested in the more short-distance-dominated cases $t = 0.2$~fm and 
$Q = 2.32$~GeV, for which the method proposed in this paper proves to be very effective. 
Similar values of $t$ and $Q$ are used in the 
analysis of the $N_f = 2$ QCD data presented in Section~\ref{sec:nf2}. 
The more infrared scales $t = 0.4$~fm and $Q = 387$~MeV are only considered in
the free-theory analysis, and they mostly serve as a comparison point.

Figure~\ref{fig:free_integrand} shows the integrand of the fourth-moment observable
$x_{\mu}^4 \mathcal G (x_{\mu})$ for all thermal lattices and their vacuum
counterparts, up to distances of $0.4$~fm.
As expected, up to around 0.2~fm the difference between the vacuum and thermal
cases is hardly noticeable. 
Also, at these short distances the cutoff effects are more important than at 
larger separations, as can be observed by comparing with the
continuum values, also displayed in the plot.
These two features motivate the use of the improved observables
defined as in Eq.~(\ref{eq:imp}) in order to achieve a better control at short distances
with the aid of fine thermal lattices. 

Before undertaking the analysis of the observables
$\mathcal I(t)$ and $\mathcal D (Q^2)$ with the method proposed
in this paper, we investigate the emergence of logarithmic cutoff effects
and compute their form explicitly. 

\subsection{A short-distance \texorpdfstring{O$(a^2 \log(1/a))$}{O(a)} cutoff effect}
\label{sect:log}

Already at the free-theory level, cutoff effects of the form
$\tilde c \: a^2 \log (1/a)$ are present in the observables $\mathcal I(t)$ and $\mathcal D (Q^2)$,
\begin{equation}
\mathcal I(t) \overset{a \to 0}{\sim} I(t) + 
\tilde c_{\mathcal I} a^2 \log (1/a) + O(a^2)
\: ,
\end{equation}
\begin{equation}
\mathcal D(Q^2) \overset{a \to 0}{\sim} D(Q^2) + 
\tilde c_{\mathcal D} a^2 \log (1/a) + O(a^2)
\: ,
\end{equation}
and analogously in the corresponding thermal quantities $\mathcal I^{\mathrm{th}}(t)$
and $\mathcal D^{\mathrm {th}} (Q^2)$. 
In the following we compute the coefficients $\tilde c_{\mathcal I}$ 
and $\tilde c_{\mathcal D}$ for the specific discretization of the 
correlation functions used in this work.

To begin with, we focus on $\mathcal I(t)$ and we analyze it in the limit $a \to 0$.
In the vacuum and in infinite volume there is no difference between projecting
to zero momentum in the directions $(x_1,x_2,x_3)$, as in Eq.~(\ref{eq:gvac}),
or in the directions $(x_0,x_1,x_2)$. Here, as in Appendix~\ref{app:free_corr},
we choose the second option, as it makes the analogy with the thermal
screening correlator very clear.
As a consequence, we will use  the vector notation 
$\vec p \equiv (p_0,p_1,p_2)$.
The thermal version of the following equations is obtained by replacing
the integral over $p_0$ with a sum over fermionic Matsubara modes, as
described in Appendix~\ref{app:free_corr},
and the coefficients $\tilde c_{\mathcal I}$ and $\tilde c_{\mathcal D}$ are the same in the vacuum and in the thermal case.
In the limit $a \to 0$, the observable $\mathcal I (t)$ can be expanded as follows\footnote{The {O}($a^2$) corrections
at the end of Eq.\ (\ref{eq:ato0}) correspond to the difference between the integral and the trapezoidal-rule based sum over $x_3$.},
\begin{equation}
\begin{split}
\mathcal I (t) = 
\int_0^t \dint x_3 \: x_3^4
\int_{-\frac{\pi}{a}}^{\frac{\pi}{a}} \frac{\dint^3 p}{(2 \pi)^3} \: 
e^{-2 p \abs{x_3}} \biggl[ \hat f_{0,0} (\hat{\vec p})\: + 
 a^2 \bigl( p^2 \hat f_{2,0}(\hat{\vec p}) + 
\abs{x_3} p^3 \hat f_{2,1}(\hat{\vec p}) \bigr) + O(a^4) \biggr] 
+ O(a^2)
\: ,
\end{split}
\label{eq:ato0}
\end{equation}
where $p \equiv \abs{\vec p}$ and $\hat f_{n,m} (\hat{\vec p})$ are
dimensionless functions of the orientation of the vector $\vec p$
($\hat{\vec p} \equiv \vec p/p$). 
The expressions of $\hat f_{00}$, $\hat f_{2,0}$ and $\hat f_{2,1}$ can be
found in Appendix~\ref{app:free_corr}.
A generic term of the expansion within square brackets in~(\ref{eq:ato0})
can be expressed as
\begin{equation}
a^n \abs{x_3}^m p^{n+m} \hat f_{n,m} (\hat{\vec p}) \: ,
\end{equation}
with $n =2,4,\dots$ and $m \geq 0$. 
The integration over $x_3$ yields
\begin{equation}
\int_0^{t} \dint x_3 \: x_3^{m+4} \: e^{-2 p x_3} =  \frac{(m+4)!}{(2 p)^{m+5}} 
-e^{-2 p t} \sum_{l=0}^{m+4} \frac{(m+4)!}{(m+4-l)!} \frac{t^{m+4-l}}{(2 p)^{l+1}}
\: .
\label{eq:z_integ}
\end{equation}
Upon integration over the Brillouin zone, the first term on the right-hand-side 
of Eq.~(\ref{eq:z_integ}) (not exponentially suppressed in $p$)
can introduce a logarithmic dependence on $a$. 
In fact, based on dimensional analysis, we observe that the term
\begin{equation}
\mathcal I (t) \supset a^n \frac{(m+4)!}{2^{m+5}}
\int_{-\frac{\pi}{a}}^{\frac{\pi}{a}} \frac{\dint^3 p}{(2\pi)^3}
\frac{\hat f_{n,m}(\hat{\vec p})}{p^{5-n}}
\label{eq:log_contrib}
\end{equation}
is proportional to $a^2 \log(1/a)$ for $n=2$, and to $a^2$ for any $n > 2$.
As a consequence, the $O(a^2)$ contribution to $\mathcal I(t)$ cannot be computed exactly
by truncating the expansion in square brackets in Eq.~(\ref{eq:ato0}) at a finite order.
Having identified the sources of logarithmic contributions,
we can compute the coefficient $\tilde c_{\mathcal I}$ by using the log-derivative
\begin{equation}
\tilde c_{\mathcal I} = \frac{1}{a} \frac{\dint}{\dint (1/a)}
\int_{-\frac{\pi}{a}}^{\frac{\pi}{a}} \frac{\dint^3 p}{(2\pi)^3} \frac{1}{p^3} 
\biggl\{ \frac{4!}{2^5} \hat f_{2,0}(\hat{\vec p})
+ \frac{5!}{2^6} \hat f_{2,1}(\hat{\vec p}) \biggl\} \biggr\rvert_{a=0} \: .
\end{equation}
The derivative of the triple integral can be computed by applying the following equation
\begin{equation}
\frac{\dint}{\dint x} \int_0^x \dint^3 p \: f(p_0,p_1,p_2) = 
\int_0^x \dint^2 p\: \bigl[ f(p_0,p_1,x) + f(p_0,x,p_1) + f(x,p_0,p_1) \bigr] \: .
\end{equation}
Using the expressions of $\hat f_{2,0}$ and $\hat f_{2,1}$ given in
Appendix~\ref{app:free_corr}, we find
\begin{equation}
\tilde c_{\mathcal I} = \frac{7 N_c}{60 \pi^2} \overset{N_c = 3}{=} \frac{7}{20 \pi^2} \: .
\label{eq:clog}
\end{equation}

\begin{table}[t]
\caption{Parameters of the continuum extrapolation of $\mathcal I^{\mathrm{th}}(t)$
and $\mathcal D^{\mathrm{th}}(Q^2)$ with the functional forms $(a)$, $(b)$, $(c)$,
$(d)$ of Eq.~(\ref{eq:fit_forms}).
The known continuum values $I^{\mathrm{th}}(t)$ and $D^{\mathrm{th}}(Q^2)$ are also reported,
together with the prefactors of the logarithmic term $\tilde c_{\mathcal I}$ (Eq.~(\ref{eq:clog}))
and $\tilde c_{\mathcal D}$ (Eq.~(\ref{eq:tildec_Adler})).
For this analysis an extended set of thermal lattices is used, with
$N_t = 16,20,24,48,60,120$ and whose temperature is fixed to $T = 246.25$~MeV.
For all fit forms, $c_0$ is in good agreement with the corresponding continuum value.
In the more infrared cases $t = 0.4$~fm and $Q = 387$~MeV the ansatz $(a)$ provides 
a good estimate of $\tilde c$, and the value of this coefficient is quite stable
with respect to introducing higher powers of $a$ in the fit ansatz.
Instead for $t = 0.2$~fm and $Q = 2.32$~GeV the role of higher-order 
discretization effects is important to obtain a relatively accurate estimate of $\tilde c$.
In particular, among the fit forms analyzed here, the ansatz $(c)$ provides the most accurate value.
}\label{tab:fit_coeff}
\centering
\begin{tabular}{ c c c c c c c c c }
\toprule
\multicolumn{9}{ c }{Free massless quarks} \\
\midrule
$t$~[fm] & $I^{\mathrm{th}}(t)~[10^{-3} \mathrm{fm}^2]$ & $\tilde c_{\mathcal I}$ &  
$c_0~[10^{-3} \mathrm{fm}^2]$ & $\tilde c$ & $c_2$ & $c_3~[\mathrm{fm}^{-1}]$ & 
$c_4~[\mathrm{fm}^{-2}]$ &  \\ 
\midrule
0.2  & 1.025 &  0.0355 & 1.026 &  0.0251 & $-0.112$ &  --       & --   & $(a)$ \\
     &       &         & 1.025 &  0.0466 & $-0.197$ & 0.522     & --   & $(b)$ \\
     &       &         & 1.025 &  0.0377 & $-0.156$ & --        & 3.70 & $(c)$ \\
     &       &         & 1.025 &  0.0292 & $-0.116$ & $-0.488$  & 7.13 & $(d)$ \\
\midrule
0.4  & 3.603 & 0.0355  & 3.603 &  0.0332 & $-0.123$ & --        & --    & $(a)$ \\
     &       &         & 3.603 &  0.0373 & $-0.139$ & 0.0992    & --    & $(b)$ \\
     &       &         & 3.603 &  0.0356 & $-0.132$ & --        & 0.703 & $(c)$ \\
     &       &         & 3.603 &  0.0352 & $-0.130$ & $-0.0231$ & 0.866 & $(d)$ \\
\midrule
$Q$~[GeV] & $D^{\mathrm{th}}(Q^2)$ & $\tilde c_{\mathcal D}[\mathrm{fm}^{-2}]$ & 
$c_0$ & $\tilde c~[\mathrm{fm}^{-2}]$ & $c_2~[\mathrm{fm}^{-2}]$ & 
$c_3~[\mathrm{fm}^{-3}]$ & $c_4~[\mathrm{fm}^{-4}]$ &  \\ 
\midrule
2.32  & 2.695  & 48.6  & 2.697  & 30.8  & $-138$   &  --  & --    & $(a)$ \\
      &        &       & 2.695  & 56.7  & $-241$   & 639  & --    & $(b)$ \\
      &        &       & 2.695  & 45.8  & $-190$   & --   & 4518  & $(c)$ \\
      &        &       & 2.695  & 52.1  & $-219$   & 369  & 1916  & $(d)$ \\
\midrule
0.387 & 0.2927 & 1.349 & 0.2927 & 1.320 & $-2.57$ & --    & --    & $(a)$ \\
      &        &       & 0.2927 & 1.368 & $-2.76$ & 1.19  & --    & $(b)$ \\
      &        &       & 0.2927 & 1.348 & $-2.67$ & --    & 8.46  & $(c)$ \\
      &        &       & 0.2927 & 1.350 & $-2.67$ & 0.108 & 7.70  & $(d)$ \\
\bottomrule
\end{tabular}
\end{table}

Moving now to the Adler function $\mathcal D(Q^2)$, we observe that
in the short-distance limit its integrand is proportional to that
of $\mathcal I (t)$
\begin{equation}
K(x_3,Q^2) \mathcal G (x_3) \overset{x_3 \to 0}{\sim}
\pi^2 Q^2 \: x_3^4 \: \mathcal G (x_3)
\: .
\end{equation}
As we saw in the above computation, the logarithmic cutoff effect comes from 
the contribution around $x_3 = 0$ to the integral over $x_3$ (see Eq.~(\ref{eq:z_integ})),
from which we conclude that
\begin{equation}
\tilde c_{\mathcal D} = \pi^2 Q^2 \: \tilde c_{\mathcal I} =
\frac{7 N_c Q^2}{60} \overset{N_c = 3}{=} \frac{7 Q^2}{20}
\: .
\label{eq:tildec_Adler}
\end{equation}

\begin{figure}[tp]
\center
\includegraphics[width = .5\textwidth]{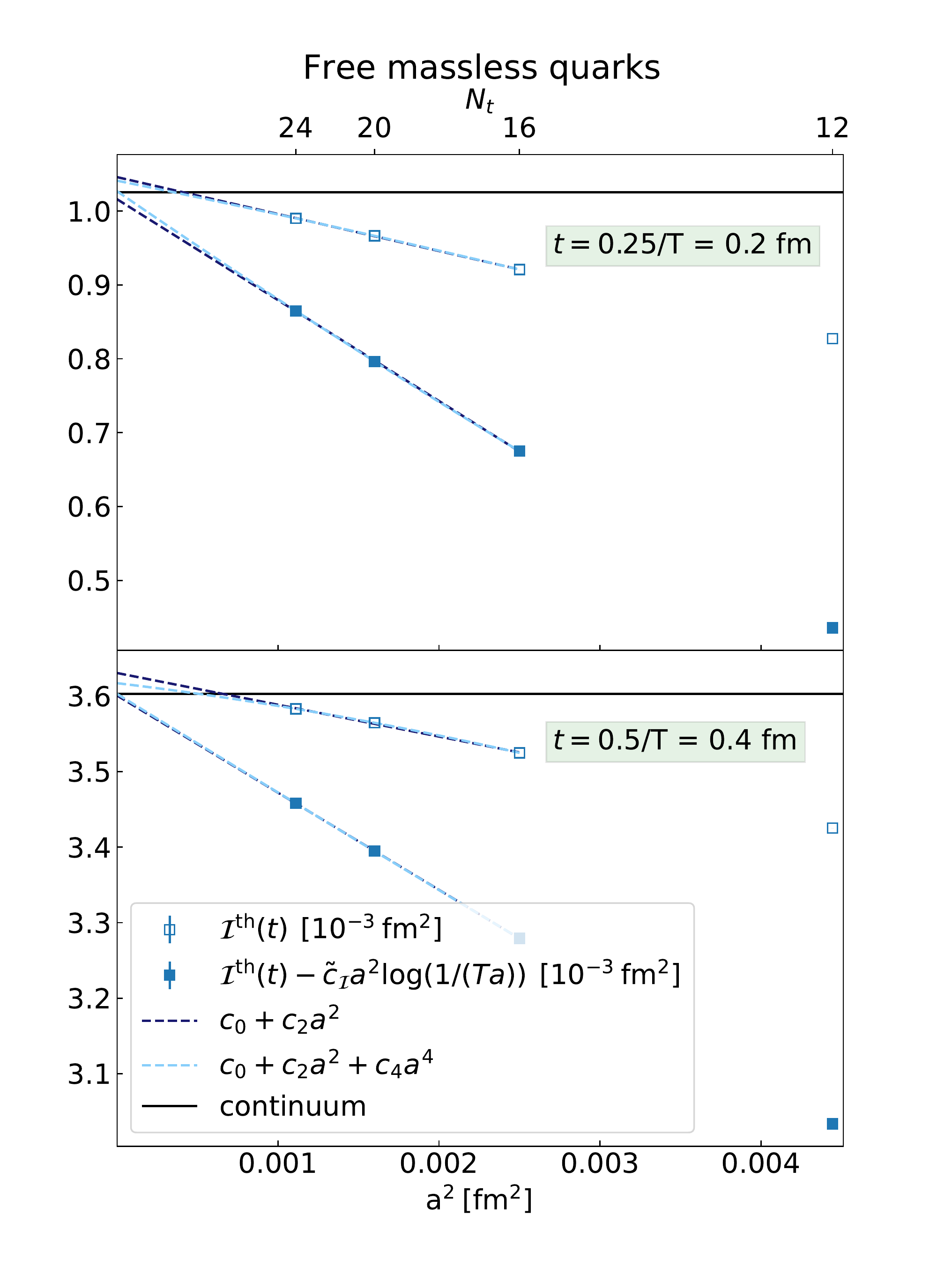}~
\includegraphics[width = .5\textwidth]{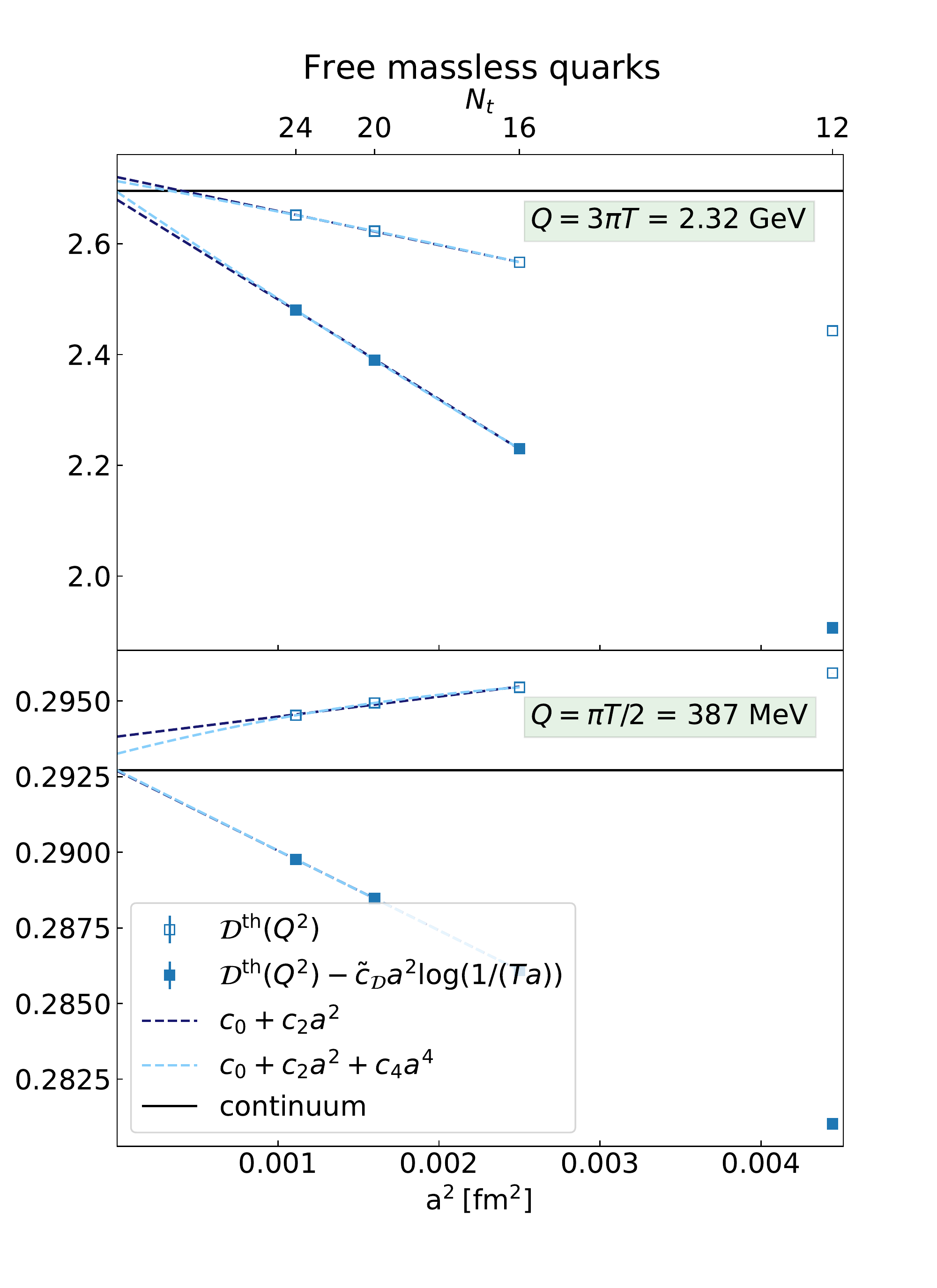}
\caption{Continuum limit of the thermal quantities $\mathcal I^{\mathrm{th}}(t)$ (left)
and $\mathcal D^{\mathrm{th}}(Q^2)$ (right). Four lattices are considered, with $N_t = 12,16,20,24$ and
whose temperature is fixed to 264.25~MeV. The coarsest lattice with $N_t = 12$ is
excluded from the fits. Polynomials of first and second degree in $a^2$ are used to fit the data.
More accurate extrapolations are obtained by
making use of the known coefficients $\tilde c_{\mathcal I}$ (\ref{eq:clog})
and $\tilde c_{\mathcal D}$ (\ref{eq:tildec_Adler}) to subtract the $O(a^2 \log(1/a))$ 
contribution from the lattice data. The relative differences between the continuum estimates shown in 
this figure and the correct continuum values are listed in Table~\ref{tab:extrap_finiteT}.
}
\label{fig:extrap_finiteT}
\end{figure}

The values of $\tilde c_{\mathcal I}$ and $\tilde c_{\mathcal D}$ 
given in Eqs.~(\ref{eq:clog}) and (\ref{eq:tildec_Adler})
can be compared to what is obtained via fits to the lattice data.
With this goal in mind, we consider a set of lattices with $N_t = 16,20,24,48,60,120$, whose 
temperature is fixed to $T = 246.25$~MeV. These include ``realistic'' lattices with
$N_t = 16,20,24$, whose lattice spacings are very close to those of the 
ensembles presented in Section~\ref{sec:nf2}, and three extremely fine lattices
with $N_t = 48,60,120$. We include the latter in order to have a better control on the 
continuum extrapolation and more flexibility with respect to the number of fit parameters.
We consider the functional forms
\begin{equation}
\begin{split}
&(a) \quad c_0 + a^2 [c_2 + \tilde c \log(1/(Ta))] \\
&(b) \quad c_0 + a^2 [c_2 + \tilde c \log(1/(Ta))] + c_3 a^3 \\
&(c) \quad c_0 + a^2 [c_2 + \tilde c \log(1/(Ta))] + c_4 a^4 \\
&(d) \quad c_0 + a^2 [c_2 + \tilde c \log(1/(Ta))] + c_3 a^3 + c_4 a^4 \\
\end{split}
\label{eq:fit_forms}
\end{equation}
and observe that the case $(c)$ leads to the best agreement with the expected value of 
$\tilde c$.
In all cases, the agreement between $c_0$ and the known continuum value is very good.
All results are reported in Table~\ref{tab:fit_coeff}.

\subsection{Continuum limit of the thermal observables}
\label{sub:cont_therm_obs}

\begin{table}[tb]
\caption{Accuracy of the continuum limit of $\mathcal I^{\mathrm{th}}(t)$ (left) and
$\mathcal D^{\mathrm{th}}(Q^2)$ (right) expressed in terms of the relative 
difference between the continuum estimate $c_0$ and the known continuum value.
Three thermal lattices are used for the extrapolation, with $N_t = 16,20,24$ and whose
temperature is set to $T=246.25$~MeV. 
The label `plain' refers to a continuum estimate obtained by fitting the plain lattice observables
$\mathcal I^{\mathrm{th}}(t)$ and $\mathcal D^{\mathrm{th}}(Q^2)$, while in the case
`subtr.' the logarithmic cutoff effects are subtracted prior to performing the fit as follows,
$\mathcal I^{\mathrm{th}}(t)-\tilde c_{\mathcal I} a^2 \log(1/(Ta))$,
$\mathcal D^{\mathrm{th}}(Q^2)-\tilde c_{\mathcal D} a^2 \log(1/(Ta))$.
The analytic values of the coefficients $\tilde c_{\mathcal I}$ and $\tilde c_{\mathcal D}$
are given in Eqs.~(\ref{eq:clog}) and (\ref{eq:tildec_Adler}) respectively.
The lattice data and the fit curves are shown in Figure~\ref{fig:extrap_finiteT}.
}\label{tab:extrap_finiteT}
\centering
\fbox{Free massless quarks} \\
\vspace{.2cm}
\begin{tabular}{c c c c c}
\toprule
$t$~[fm] & \multicolumn{2}{c}{$\abs{c_0 - I^{\mathrm{th}}(t)}/I^{\mathrm{th}}(t)$} & ansatz  \\
\midrule
0.2 & 2\%   & 0.9\%    & $c_0 + c_2 a^2$             \\
    & 2\% & 0.2 \%     & $c_0 + c_2 a^2 + c_4 a^4$   \\
0.4 & 0.8\% & 0.06 \%  & $c_0 + c_2 a^2$             \\
    & 0.4\% & < 0.01\% & $c_0 + c_2 a^2 + c_4 a^4$   \\
\midrule
         & plain & subtr. &   \\
\bottomrule
\end{tabular}
\hspace{.2cm}
\begin{tabular}{c c c c c}
\toprule
$Q$~[GeV] & \multicolumn{2}{c}{$\abs{c_0 - D^{\mathrm{th}}(Q^2)}/D^{\mathrm{th}}(Q^2)$} & ansatz  \\
\midrule
2.32  & 0.9\% & 0.6\%    & $c_0 + c_2 a^2$             \\
      & 0.7\% & 0.06 \%  & $c_0 + c_2 a^2 + c_4 a^4$   \\
0.387 & 0.4\% & < 0.01\% & $c_0 + c_2 a^2$             \\
      & 0.2\% & < 0.01\% & $c_0 + c_2 a^2 + c_4 a^4$   \\
\midrule
         & plain & subtr. &   \\
\bottomrule
\end{tabular}
\end{table}

As a first step toward improved vacuum observables defined as in Eq.~(\ref{eq:imp}),
we compute continuum estimates of the thermal quantities $\mathcal I^{\mathrm{th}}(t)$ and 
$\mathcal D^{\mathrm{th}}(Q^2)$.
We consider a set of four lattices with $N_t = 12,16,20,24$, whose temperature
is set to 246.25~MeV and whose lattice spacings are very similar to the ones
of the $N_f = 2$ thermal ensembles, as discussed in Section~\ref{sect:latt_form}.
In all practical cases, we find it expedient to exclude the coarser lattice
with $N_t = 12$ from the continuum extrapolation.

\begin{figure}[t]
\center
\includegraphics[width = .5\textwidth]{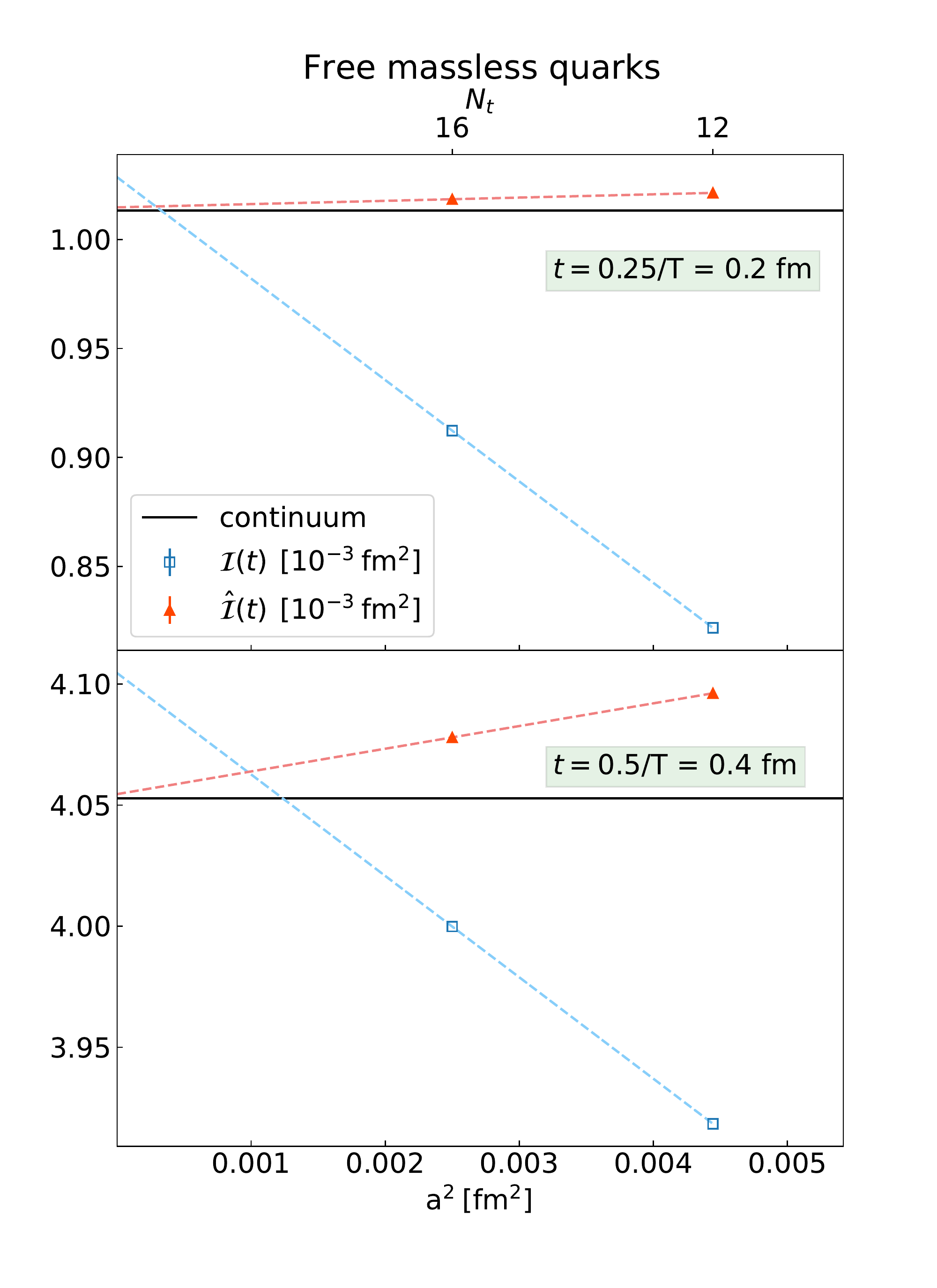}~
\includegraphics[width = .5\textwidth]{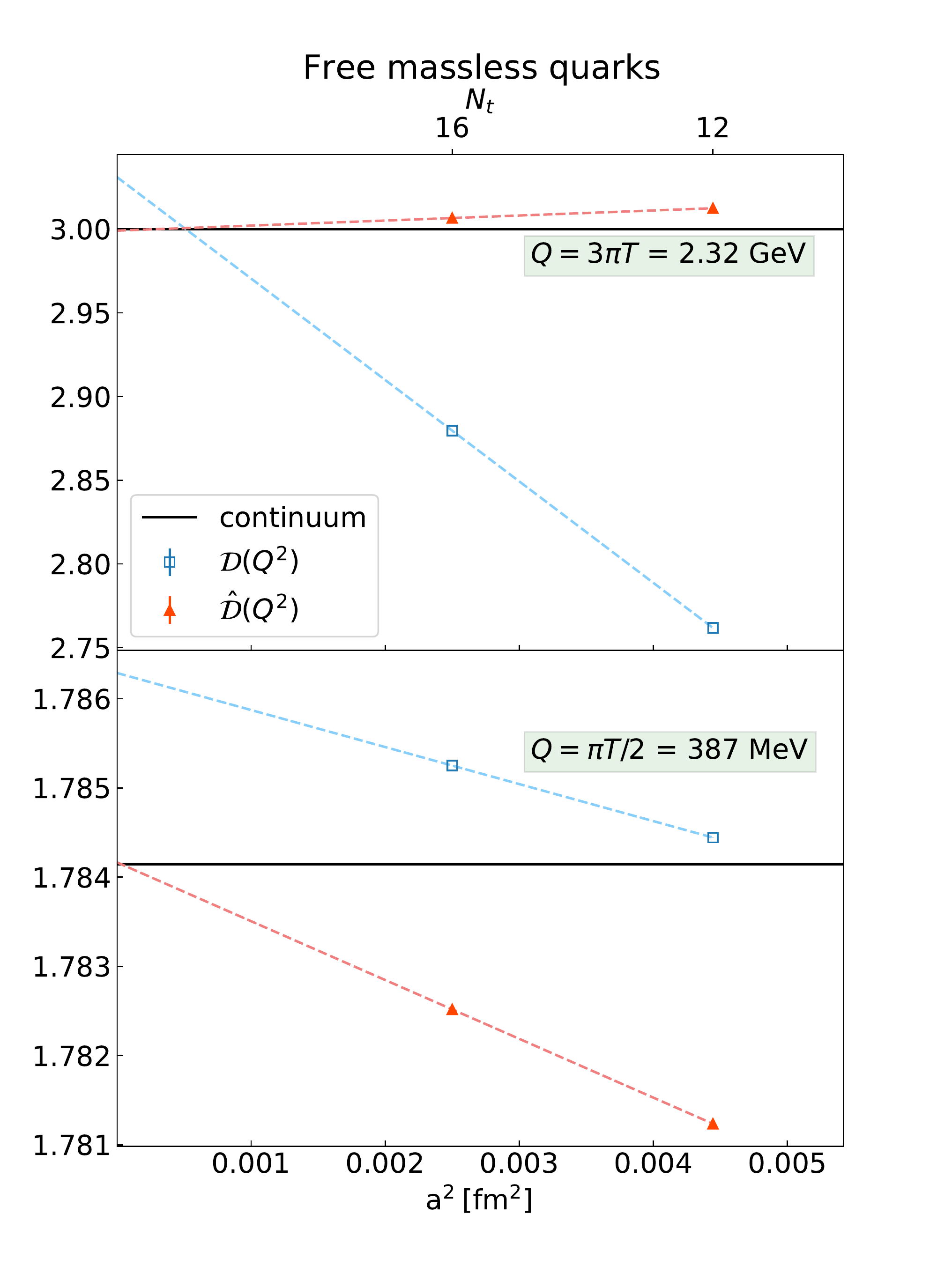}
\vspace{-0.6cm}
\caption{Continuum extrapolation of the vacuum lattice observables $\mathcal I(t)$ (left) and 
$\mathcal D(Q^2)$ (right) and of their improved versions $\hat{\mathcal I}(t)$ and $\hat{\mathcal D}(Q^2)$.
The observables are evaluated on two lattices whose lattice spacings $a \approx \{0.07,0.05\}$~fm 
are equal to those of the $N_t = 12,16$ thermal lattices. The fit ansatz is linear in $a^2$.
The accuracy of the resulting continuum estimates is reported in Table~\ref{tab:extrap_vac}.
}
\label{fig:extrap_vac}
\end{figure}

We observe that a careful treatment of the logarithmic cutoff effects can significantly 
improve the accuracy of the continuum limit. 
In order to illustrate this point, we compare the outcome of two different approaches.
The first is to simply ignore the presence of logarithmic cutoff effects and to fit
the lattice data with polynomials in $a^2$.
The second is to subtract the logarithmic contributions making use of the known prefactors
$\tilde c_{\mathcal I}$ and $\tilde c_{\mathcal D}$ (see Section~\ref{sect:log}) before
fitting polynomially in $a^2$.
This second approach proves to be very effective, however it is somewhat specific to the free 
theory. In the interacting case, it remains to be seen whether the $a^2 \log(1/a)$ term
receives significant, non-analytic in $a$ corrections.
Another possible strategy is to fit the lattice data with the ansatz 
$c_0 + c_2 a^2 + \tilde c a^2 \log(1/(Ta))$.
In this case we find that the accuracy of the continuum estimates is comparable with the results obtained by subtracting 
the logarithmic term and then fitting linearly in $a^2$, which are given in Table~\ref{tab:extrap_finiteT}.

In the left panel of Figure~\ref{fig:extrap_finiteT} two sets of lattice data are shown,
one corresponds to the plain thermal observable $\mathcal I^{\mathrm{th}}(t)$ (open symbols) and
the other represents $\mathcal I^{\mathrm{th}}(t) - \tilde c_{\mathcal I} a^2 \log(1/(Ta))$
(filled symbols). The value of the coefficient $\tilde c_{\mathcal I}$ is 
given in Eq.~(\ref{eq:clog}).
Four different continuum estimates are obtained by fitting these data sets with two polynomial forms,
one linear in $a^2$ and one quadratic in the same variable.
The discrepancy between these estimates and the known continuum value is reported on the left-hand side
of Table~\ref{tab:extrap_finiteT}.
The accuracy of the continuum limit is significantly improved by the subtraction of the logarithmic 
term and the inclusion of the $O(a^4)$ term also plays an important role.
For example, for $t = 0.2$~fm a naive polynomial fit of the lattice data gives a
discrepancy with the correct continuum value of around $2\%$, 
which is reduced to $0.2\%$ by subtracting the logarithmic cutoff effects and
fitting the resulting lattice points with a second-degree polynomial in $a^2$.
As final continuum estimates we choose the most accurate results
\begin{equation}
I^{\mathrm{th}}_{\mathrm{extrap.}} (t = 0.2~\mathrm{fm}) =  1.027 \times 10^{-3}~\mathrm{fm}^2 \: ,
\quad
I^{\mathrm{th}}_{\mathrm{extrap.}} (t = 0.4~\mathrm{fm}) =  3.603 \times 10^{-3}~\mathrm{fm}^2 \: .
\label{eq:cont_mom}
\end{equation}

A similar analysis of the thermal Adler function $\mathcal D^{\mathrm{th}}(Q^2)$ can be found on the right
panel of Figure~\ref{fig:extrap_finiteT} and on the right-hand side of Table~\ref{tab:extrap_finiteT}.
Also for this observable the gain in accuracy due to subtracting the logarithmic cutoff effects
is considerable.
For example, for $Q = 2.32$~GeV the continuum estimate obtained by fitting $\mathcal D^{\mathrm{th}}(Q^2)$
 with a second-degree polynomial in $a^2$ differs from the correct continuum value by $0.7\%$, while 
fitting $\mathcal D^{\mathrm{th}}(Q^2) - \tilde c_{\mathcal D} a^2 \log(1/(Ta))$
with the same ansatz reduces the discrepancy to $0.06\%$. 
The value of $\tilde c_{\mathcal D}$ is given in Eq.~(\ref{eq:tildec_Adler}).
As final continuum estimates of the thermal Adler function we choose the most accurate values
\begin{equation}
D^{\mathrm{th}}_{\mathrm{extrap.}} (Q = 2.32~\mathrm{GeV}) =  2.694  \: ,
\quad
D^{\mathrm{th}}_{\mathrm{extrap.}} (Q = 387~\mathrm{MeV}) =  0.2927 \: .
\label{eq:cont_Ad}
\end{equation}

\begin{table}[h!]
\caption{Accuracy of the continuum estimates obtained by fitting with the ansatz
$c_0 + c_2 a^2$ the lattice observables $\mathcal I(t)$, $\mathcal D(Q^2)$ and their
improved versions $\hat{\mathcal I} (t)$, $\hat{\mathcal D} (Q^2)$ defined in 
Eqs.~(\ref{eq:Iimpr}) and (\ref{eq:Dimpr}).
The accuracy is expressed in terms of the relative difference with the known continuum
values $I(t)$ and $D(Q^2)$.
The lattice data and the fit curves are shown in Figure~\ref{fig:extrap_vac}.
}\label{tab:extrap_vac}
\centering
\fbox{Free massless quarks} \\
\vspace{.1cm}
\begin{tabular}{c c c c}
\toprule
$t$~[fm] & \multicolumn{2}{c}{$\abs{c_0 - I(t)}/I(t)$} \\
\midrule
0.2 & 2\%   & 0.2\%     \\
0.4 & 1\% & 0.04 \%   \\
\midrule
         & $\mathcal I$ & $\hat{\mathcal I}$ &   \\
\bottomrule
\end{tabular}
\hspace{.5cm}
\begin{tabular}{c c c c}
\toprule
$Q$~[GeV] & \multicolumn{2}{c}{$\abs{c_0 - D(Q^2)}/D(Q^2)$}  \\
\midrule
2.32  & 1\% & 0.03\%      \\
0.387 & 0.1\% & < 0.01\%   \\
\midrule
         & $\mathcal D$ & $\hat{\mathcal D}$ &   \\
\bottomrule
\end{tabular}
\end{table}

\subsection{Continuum limit of the improved vacuum observables}
\label{sub:impr_vacuum}

With the continuum estimates of Eqs.~(\ref{eq:cont_mom}) and (\ref{eq:cont_Ad}), 
we build the improved vacuum observables
\begin{equation}
\hat {\mathcal I}(t) = 
I^{\mathrm{th}}_{\mathrm{extrap.}} (t) +
[\mathcal I (t) - \mathcal I^{\mathrm{th}} (t) ]
\: ,
\label{eq:Iimpr}
\end{equation}
\begin{equation}
\hat {\mathcal D} (Q^2) = 
D^{\mathrm{th}}_{\mathrm{extrap.}} (Q^2) +
[\mathcal D (Q^2) - \mathcal D^{\mathrm{th}} (Q^2) ]
\: .
\label{eq:Dimpr}
\end{equation}
As in the case of the $N_f = 2$ QCD ensembles, we evaluate the observables on
two zero-temperature lattices, whose lattice spacings $a \approx \{0.07,0.05\}$~fm are the
same as those of the $N_t = 12,16$ thermal lattices.
We obtain continuum estimates by fitting linearly in $a^2$ the lattice observables $\mathcal I(t)$,
$\mathcal D(Q^2)$ and their improved versions defined in Eqs.~(\ref{eq:Iimpr}) and (\ref{eq:Dimpr}).
The lattice data and the fit curves are shown in Figure~\ref{fig:extrap_vac}, while the accuracy of the 
resulting continuum estimates, given in terms of the relative difference with the correct continuum value, 
is reported in Table~\ref{tab:extrap_vac}.
For the case of $\mathcal I(t=0.2\,{\rm fm})$,  one order of magnitude in accuracy is gained by using the improved lattice 
observables introduced in this paper.

In all cases considered here, the advantage of using the improved observables $\hat{\mathcal I}(t)$,
$\hat{\mathcal D}(Q^2)$ is rather clear as far as the accuracy of the resulting continuum estimates is concerned.
For the more ultraviolet scales $t = 0.2$~fm and $Q = 2.32$~GeV there is the further
benefit of a significant reduction of the cutoff effects at finite lattice spacing, as compared to
the plain lattice observables $\mathcal I(t)$, $\mathcal D(Q^2)$.
As a final remark, we repeat that the logarithmic discretization effects proportional to $a^2\log(1/a)$ cancel in 
$\hat{\mathcal I}(t)$ and $\hat{\mathcal D}(Q^2)$, due to the subtraction between the vacuum and thermal lattice observables.

\section{Non-perturbative test in \texorpdfstring{$\Nf=2$}{Nf=2} QCD\la{sec:nf2}}
In this section, we perform a non-perturbative numerical study of the same
observables investigated in the free theory in the previous section, namely the
(truncated) fourth moment of the current correlator and the Adler function at large
virtuality.
We make use of the vacuum CLS ensembles with $\Nf=2$ non-perturbatively
O($a$)-improved Wilson fermions and the Wilson gauge action with two lattice
spacings of $a\approx 0.049$~fm and $a=0.0658$~fm.
Our study is performed at a fixed, common mass of the up and down quarks.
For the (zero-temperature) pion mass, we quote the values 268(3)\,MeV and 269(3)\,MeV
respectively for ensembles F7 and O7~\cite{Engel:2014cka}.
The improved estimators were computed using ensembles on the same line of
constant physics set by the physical volume $L$ and quark mass $m_\mathrm{q}$
with a temperature of  $T=250$~MeV.

The aspect ratio for the finite-temperature ensembles was set to $L_0/L=1/4$,
and for the vacuum ensembles to $L_0/L=2$.
The thermal ensembles were recently used in a study of the photon emissivity of the quark-gluon plasma~\cite{Ce:2020tmx}.
Two of them have common bare parameters with the vacuum ensembles, while two
additional ensembles with lattice spacings down to $a\simeq
0.033$~fm allow the continuum limit of the thermal observable
to be obtained with reduced uncertainty.
Further details on the ensembles are collected in Table~\ref{tab:nf2}.%
\footnote{It is worth recording that the configurations of the $24\times 96^3$ X7 ensemble
could be generated at the cost of 1.9 million core hours on a compute cluster
equipped with Intel Skylake processors and a 50\,GBit/s Omnipath network.}
The scale was set for the F7 ensemble taking the lattice spacing from
ref.~\cite{Fritzsch:2012wq}, and assuming a perfect line of constant physics
with a ratio of lattice spacings of 3/4 between O7 and F7.
The assumed ratio of lattice spacings is consistent at the one-sigma level with the values of the
lattice spacings given in~\cite{Fritzsch:2012wq}, as well as with those of  Ref.~\cite{Engel:2014cka},
in which they are quoted with a 0.9\% precision.
Note that in contrast to the free massless theory, in the present case the current is
not fully O($a$)-improved as the improvement coefficients are not known
non-perturbatively.
Nevertheless, at high temperature, the O($a$) discretization effects due to the missing
current improvement terms should be proportional to the quark mass O($am_\mathrm{q})$, owing to the restoration of
chiral symmetry in the massless theory~\cite{DallaBrida:2020gux,Brandt:2017vgl}.

\input{table_nf2}

The integrands for the two observables considered here are displayed
in Figure~\ref{fig:nf2_compare} for the vacuum and thermal O7
ensembles.
For both observables $\mathcal O=\mathcal I, \mathcal D$, we employ linear or
quadratic fit ans\"atze and,  for the thermal observable, additionally an ansatz
where the logarithm is included using the coefficient determined at leading
order in the previous section
\begin{align}
    (\mathrm{L}) \quad & c_0 + c_1 a,\\
    (\mathrm{Q}) \quad & c_0 + c_2 a^2,\\
    (\mathrm{Q + log}) \quad & c_0 + a^2[c_2 + \tilde c_\mathcal{O}\log(L_0/a)].
    \label{eq:nf2_ansatz}
\end{align}
For the thermal observables we also use the leading-order result of the
previous section to implement an additive perturbative improvement according to
\begin{align}
    \mathring{\mathcal O}^\mathrm{th}(t,a) &= \mathcal O^\mathrm{th}(t,a)
    - \Big[\mathcal O^\mathrm{th}(t,a) - O^\mathrm{th}(t)\Big]_\mathrm{LO}.
    \label{eq:tli}
\end{align}
We now discuss these two observables in turn.

\input{table_nf2_fits}

\subsection{The short-distance contribution to \texorpdfstring{$\Pi'(Q^2=0)$}{Pi'(Q2=0)}}
\label{sub:shortdist_Piprime}

\begin{figure}[tp]
    \centerline{
    \includegraphics[scale=0.65]{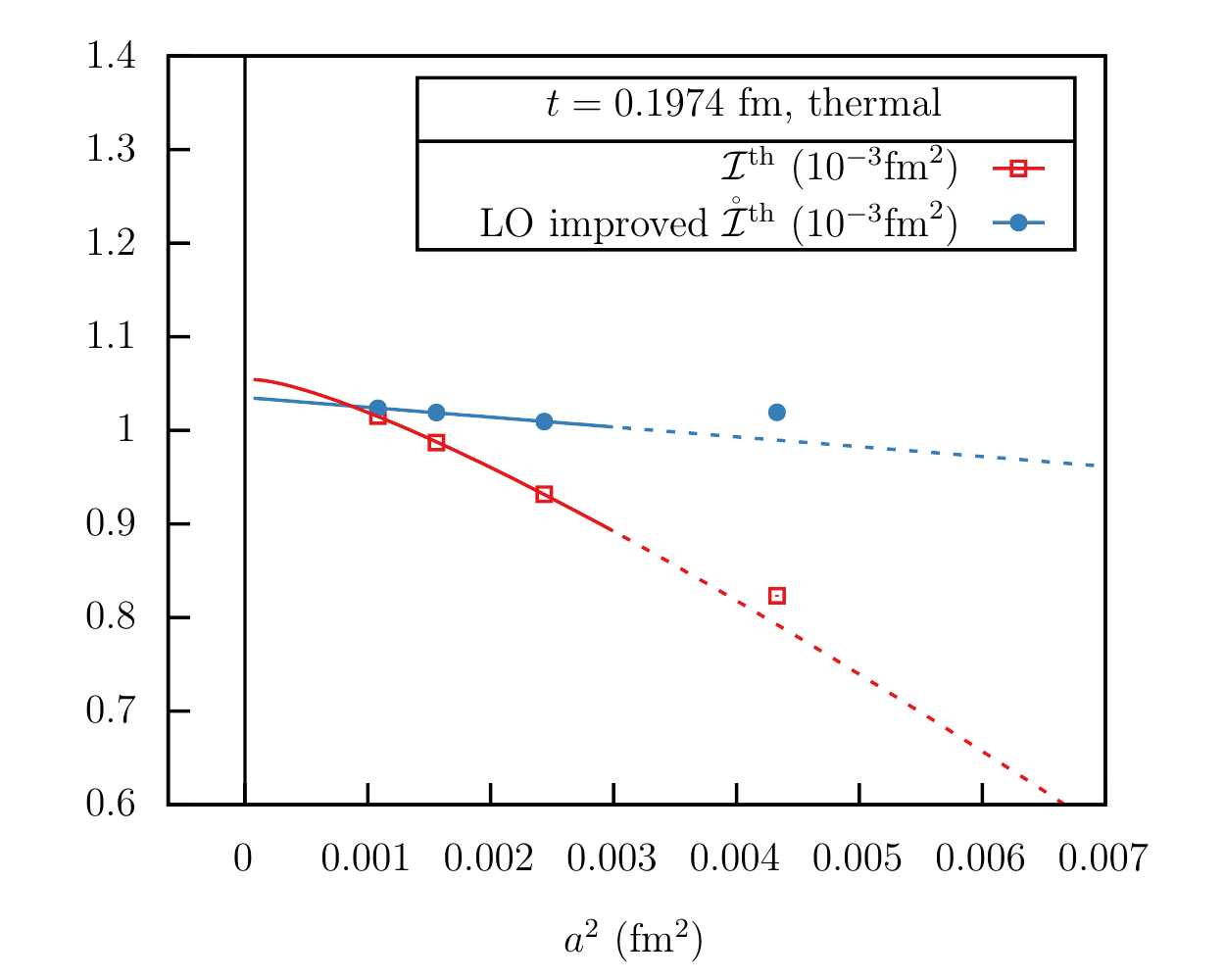}
    \includegraphics[scale=0.65]{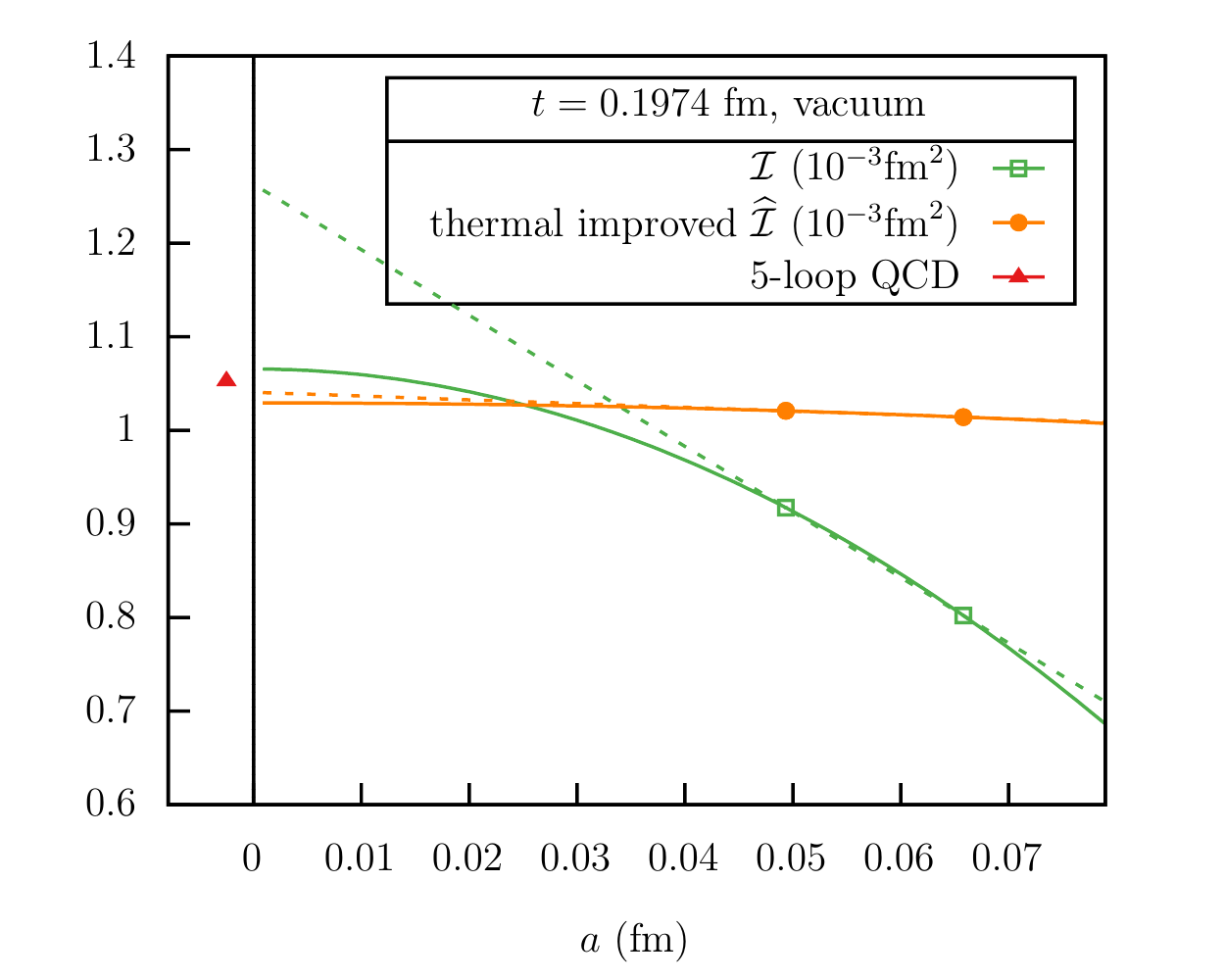}
    }
    \caption{
      Left: Continuum limit for the truncated fourth moment of the thermal correlator, $I^{\rm th}(t=0.1974{\rm \,fm})$.
      The open symbols represent the uncorrected observable, the corresponding curve showing the (Q+log) fit
      to the data points at the three finest lattice spacings.
      The filled symbols, the three leftmost of which are fitted linearly in $a^2$, represent the leading-order improved observable.
    Right: Continuum limit for the truncated fourth moment of the vacuum correlator, $I(t=0.1974{\rm \,fm})$.
    The open symbols represent the straightforward estimator ${\cal I}(t,a)$, while the 
    filled ones represent the estimator $\widehat{\cal I}(t,a)$ of Eq.\ (\ref{eq:imp}).
    Both data sets are fitted linearly either in $a$ or in $a^2$.
		We also estimated the observable $I(t)$ using the
		perturbative five-loop vacuum spectral function, following
		Refs.~\cite{Burnier:2012ts,Baikov:2008jh}, depicted with the red point.
}
    \label{fig:nf2_amu}
\end{figure}

First we examine the estimate of the continuum limit of the truncated fourth
moment of the thermal correlator Eq.~\eqref{eq:integral}, which is required to
compute the improved estimator Eq.~\eqref{eq:imp}.
In the left panel of Figure~\ref{fig:nf2_amu}, the integral up to $t=L_0/4 = 0.1974$\,fm is
shown as a function of the lattice spacing, for the thermal observable.
While the (Q+log) fit provides a satisfactory description of the data when the coarsest lattice
spacing is omitted,
the perturbatively-improved observable has a much flatter behaviour toward the
continuum.
The fit results are given in Table~\ref{tab:nf2}.

The continuum estimate from the leading-order improved extrapolation is used to
define the improved estimator of Eq.~\eqref{eq:imp}, which is shown in
the right panel of Figure~\ref{fig:nf2_amu}.
The original data are displayed as open symbols; they exhibit a large cutoff
effect.  For illustration, two fit ans\"atze are employed, purely linear or purely quadratic
in the lattice spacing. The latter may seem more plausible here, given the short-distance
nature of the observable. Nevertheless, the severity of the cutoff effect leads to an unsatisfactory
control of the continuum limit using only vacuum correlators with the available
state-of-the-art lattice spacings.
On the other hand, the thermal-improved estimator depicted with filled symbols
shows an almost flat continuum limit, which suggests the subtraction of the
thermal contribution also removes a significant amount of the cutoff effects,
as expected. In this case, the continuum result is much less sensitive to the choice
of fit ansatz for taking the continuum limit.

In order to quote a continuum estimate for the thermal-improved observable for
illustration, we choose to use the continuum estimate of the thermal observable
obtained with the LO-improvement, and for the correction the mean of the
continuum estimates obtained with the linear and quadratic fits to arrive at
\begin{align}
  I(t) &= 1.035(9)_\mathrm{stat}(19)_\mathrm{cont} \times 10^{-3}\;\mathrm{fm}^2,
  \qquad\quad  t=0.1974\,\mathrm{fm}.
    \label{eq:final_moment}
\end{align}
The second, systematic error associated with taking the continuum limit is estimated as the quadrature sum of (a)
  the difference between the (Q+log) extrapolation of ${\cal I}^{\rm th}$ and
  the (Q) extrapolation of $\mathring{\cal I}^{\rm th}$,
  and (b) half the difference between the
linear and quadratic continuum fits of the correction term.

Finally, we compute the same observable in $\Nf=2$ massless perturbation theory based on the spectral representation~\cite{Bernecker:2011gh}
\begin{equation}
I(t) = \int_{2m_\pi}^{\infty} \dint \omega\, \omega^2\, \rho(\omega^2) \,\frac{\mathrm{d}^4}{\mathrm{d}\omega^4}\left(\frac{1-e^{-\omega t}}{\omega}\right),
\end{equation}
where $\rho(\omega^2)$ is the five-loop massless vacuum spectral function\footnote{Our convention for the normalization of the
spectral functions is such that for the electromagnetic current correlator, $\rho(s)= R(s)/(12\pi^2)$, with $R(s)$
the ratio of cross-sections for $e^+e^- \to$ hadrons over $e^+e^- \to \mu^+\mu^-$.}~\cite{Burnier:2012ts,Baikov:2008jh}.
We use the renormalization scale $\mu=2.4$ GeV and take $\Lambda_{\overline{\rm MS}}^{(2)}$
from the FLAG report~\cite{Aoki:2019cca}.
The result obtained is $I_\mathrm{pert}(t=0.1974\mathrm{fm})=1.059(-6)(1)\times
10^{-3}\mathrm{fm}^2$, which is depicted with the filled red point to the left
in Figure~\ref{fig:nf2_amu}.
The errors are estimated using the uncertainty in $\Lambda_{\overline{\rm MS}}^{(2)}$.
We find reasonable agreement between our final estimate and
perturbation theory, even though we have not investigated the systematic effects associated with finite quark masses
and residual non-perturbative effects, which might become relevant at the quoted level of precision.
Also, we have not included the scale-setting uncertainty in Eq.\ (\ref{eq:final_moment});
the relative scale-setting uncertainty of $I(t)$ is mainly that of $t^2/{\rm fm}^2$, i.e.\ about 2\%,
whereas $I(t)/t^2$ only depends weakly on $t$ around $t=0.2\,$fm.

\begin{figure}[tp]
    \centerline{
    \includegraphics[scale=0.65]{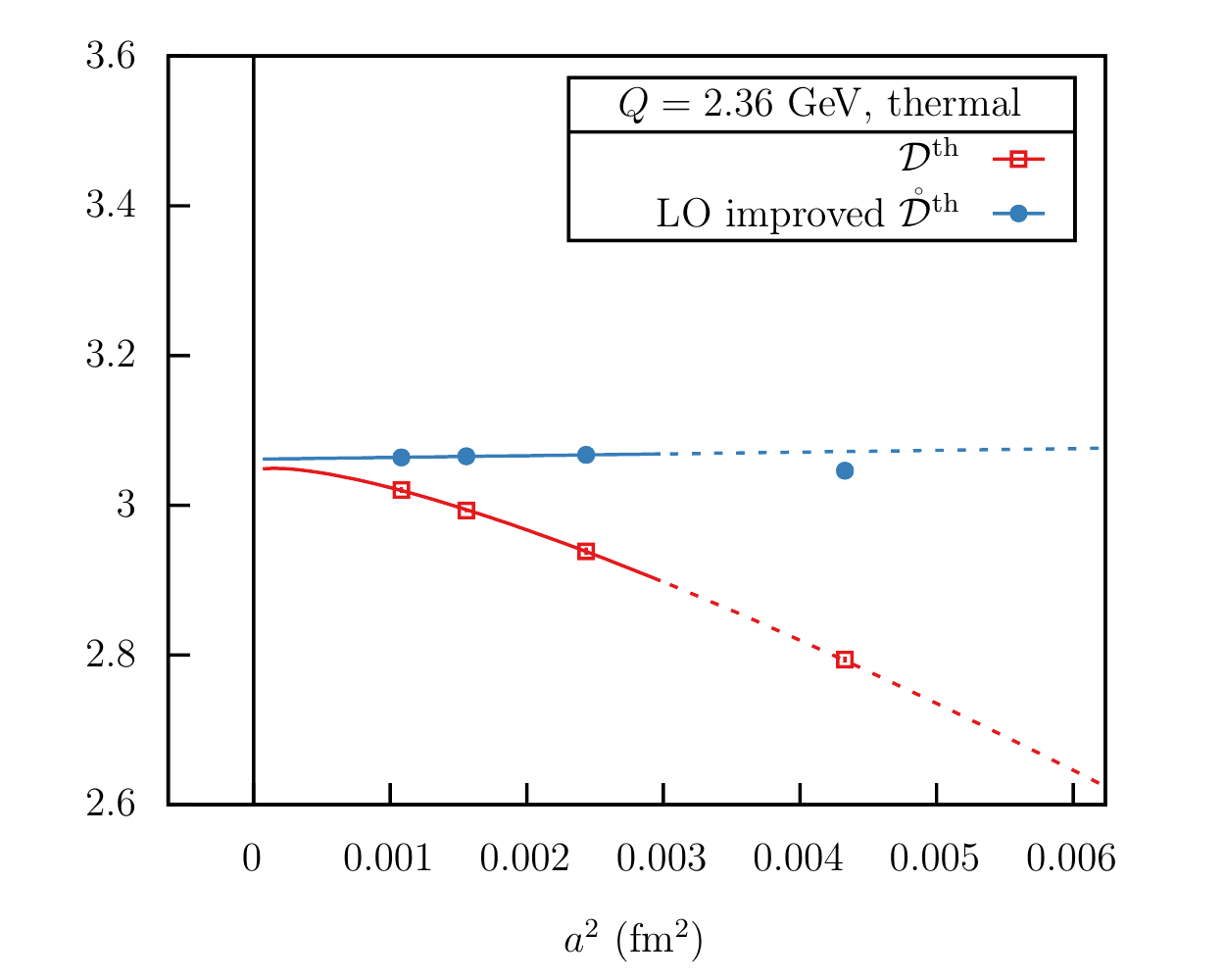}
    \includegraphics[scale=0.65]{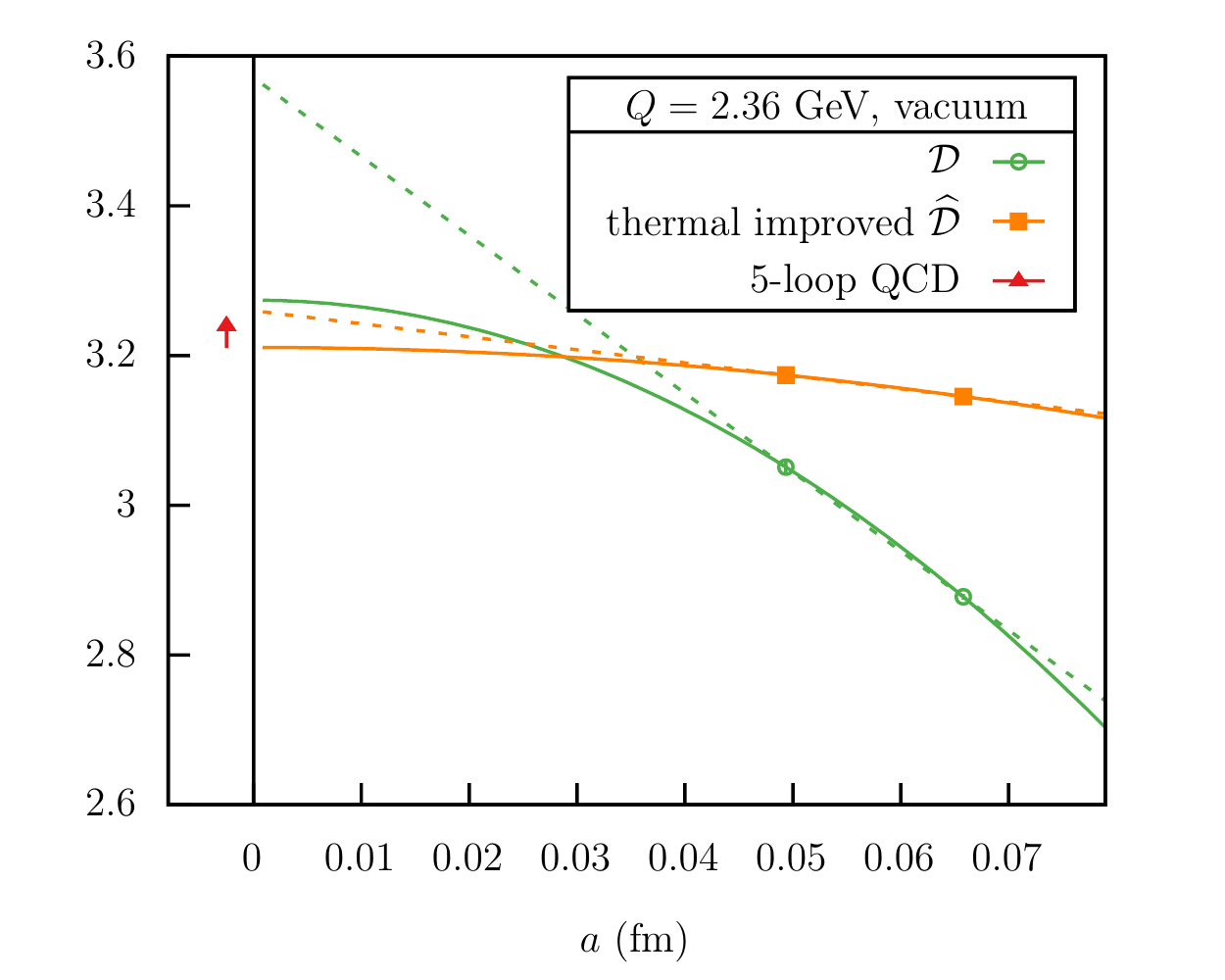}
    }
    \caption{
        Continuum limit for the `thermal' Adler function $D^{\rm th}(Q^2)$ (left) and the corresponding vacuum
        observable $D(Q^2)$ (right) for a large virtuality $Q=2.36$\,GeV.
        The various data symbols and curves are entirely analogous to those in figure~\ref{fig:nf2_amu}.
    }
    \label{fig:nf2_adler}
\end{figure}

\subsection{The Adler function at large \texorpdfstring{$Q^2$}{Q2}}
\label{sub:adler_largeq2}

The continuum limit of the thermal Adler function for a large value of the
virtuality $Q= 3\pi T = 2.36$\,GeV is shown in Figure~\ref{fig:nf2_adler}.
As in the previous case, the leading-order improved observable shows better
scaling to the continuum limit, when the coarsest lattice spacing is not
included.
The original and thermal-improved estimator
are shown in the right-hand panel of Figure~\ref{fig:nf2_adler}, which
likewise strongly suggests the suppression of lattice artifacts in the
difference.
Once again, the difficulty of performing a controlled continuum limit purely based
on the vacuum correlators at large virtualities is apparent even with fine lattice
spacings available.
For illustration, we quote an estimate for the improved observable where the
central value and systematic error are determined in the same way as in the
subsection before
\begin{align}
  D(Q^2) &= 3.24(4)_\mathrm{stat}(3)_\mathrm{cont},
  \qquad \qquad Q=2.36\,{\rm GeV}.
    \label{eq:final_adler}
\end{align}
For the five-loop perturbative result, using
\be
D(Q^2) = 12\pi^2\,Q^2 \int_{4m_\pi^2} ^\infty \dint\omega^2\,\frac{\rho(\omega^2)}{(\omega^2+Q^2)^2} 
\ee
and the renormalization scale $\mu=Q$, we found $D_\mathrm{pert}(Q^2) = 3.24(-3)(1)$.
Clearly, the perturbative prediction is in good agreement with our non-perturbative estimate obtained
with the help of thermal correlators computed at very fine lattice spacings.

\section{Summary and outlook\la{sec:concl}}

We have shown that a better control over the continuum limit for
short-distance dominated integrals over the vector correlator can be
achieved by using thermal screening correlators at particularly fine
lattice spacings, and then correcting for the difference.  For the
Adler function at a momentum $Q=2.36$\, GeV, we estimate that we
achieved a reduction of the systematic error due to the continuum
limit by a factor of four, relative to the continuum limit based on
the available vacuum correlators only. This is significant, since
taking the continuum limit is responsible for one of the leading
systematic errors on this quantity. The cost of generating the thermal
ensembles at small lattice spacings is modest in comparison to the
cost of the vacuum ensembles. The method can also be applied to the
charm contribution, which is even more short-range and therefore more
susceptible to large cutoff effects.

For the anomalous magnetic moment of the muon, we find that, in the
time-momentum representation, performing a `naive' linear
extrapolation of the $t\leq0.2$\,fm contribution using lattice
spacings down to 0.049\,fm leads to an overestimate of about three
percent as compared to our best estimate based on lattice spacings
down to 0.033\,fm.  Since the muon $(g-2)$ is an observable which is
much more infrared-weighted, the slightly inaccurate continuum
extrapolation of the short-distance contribution in the $(u,d,s)$
quark sector only amounts to a difference of about $0.4\times
10^{-10}$ on this quantity, which is an order of magnitude smaller
than the precision of the current most precise
estimates~\cite{Aoyama:2020ynm,Borsanyi:2020mff}.  Nevertheless, we
have seen clear evidence that performing additively tree-level improvement 
reduces the lattice artifacts in this short-distance regime and we
thus recommend its use in future calculations of the leading hadronic
contribution to $(g-2)_\mu$.

One may wonder, is it possible to calculate the hadronic contribution
to the running of the QED coupling up to the $Z$-boson mass in lattice
QCD with controlled errors?  The methods and tests presented in this
paper strongly suggest that it is indeed possible, and we now sketch a
promising strategy. Let $\Delta_2\Pi(Q)\equiv \Pi(Q^2) - \Pi(Q^2/4)$
be the difference of vacuum polarisations corresponding to the running
of $\alpha$ over the momentum interval $Q/2$ to $Q$. Such an
observable is very similar to the Adler function $D(Q^2)$, in that it
is dominated by Euclidean distances of order $1/Q$.  What we have seen
above leads us to conclude that $\Delta_2\Pi(Q^2)$ for a large $Q\gg
1$\,GeV can be computed at the one-percent level at a temperature
$T\approx Q/(8\pi)$ in terms of the screening vector correlator; we
may write this quantity $\Delta_2^{\rm th}\Pi(Q;T)$. The difference
$\Delta_2^{\rm th}\Pi(Q;T/2) - \Delta_2^{\rm th}\Pi(Q;T) $ can be
evaluated by performing simulations at temperatures differing by a
factor of two with common lattice spacing. See Appendix~\ref{sec:Delta2Pi} for
an encouraging study at leading order in perturbation theory. Parametrically, the
difference between $\Delta_2\Pi(Q)$ and  $\Delta_2^{\rm th}\Pi(Q;T)$ is
of order $(\pi T/Q)^4$, and thus $\Delta_2^{\rm th}\Pi(Q;T/2) -
\Delta_2^{\rm th}\Pi(Q;T) $ already provides a sufficiently good
estimate for that difference. The latter can also be estimated using
perturbation theory, since a high value of $Q$ is tied to a high value
of $T$. Thus by varying the temperature by factors of two, we can map
out the running of $\alpha$ by factors of two in $Q$ up to the $Z$-boson
mass. Such a program can be carried out on existing computing
platforms, albeit at a significant investment, since realizing the
double hierarchy $ \pi T\ll Q\ll \pi/a$ typically requires the use of
lattices of size $48\times 192^3$. If one resorts to such fine
lattices, it is probably mandatory to address the freezing of the
topological charge, for instance by using open boundary conditions in
the $x_3$-direction~\cite{Luscher:2011kk,Florio:2019nte}.

\acknowledgments
This work was supported by the
European Research Council (ERC) under the European Union’s Horizon
2020 research and innovation program through Grant Agreement
No.\ 771971-SIMDAMA, as well as by the Deutsche
Forschungsgemeinschaft (DFG, German Research Foundation) 
through the Cluster of Excellence “Precision Physics, Fundamental
Interactions and Structure of Matter” (PRISMA+ EXC 2118/1) funded by
the DFG within the German Excellence strategy (Project ID 39083149).
The work of M.C.\ is supported by the European Union's Horizon
2020 research and innovation program
under the Marie Sk\l{}odowska-Curie Grant Agreement No.\ 843134-multiQCD.
T.H. is supported by UK STFC CG ST/P000630/1.
The generation of gauge configurations as well as
the computation of correlators was performed on the Clover and Himster2 platforms
at Helmholtz-Institut Mainz and on Mogon II at Johannes Gutenberg University Mainz.
We have also benefitted from computing resources at Forschungszentrum J\"ulich allocated under NIC project HMZ21.

\appendix

\section{Derivation of the OPE for the vector correlator}
\label{app:OPE}

In this appendix we use the operator-product expansion (OPE) as a tool to study the static screening correlator at short distance.
To leading order in the expansion, the vacuum and thermal correlation functions are identical.
The following term, linear in the distance $\abs{x_3}$, is in general non-vanishing for the thermal correlator.
Here we ask ourselves if a suitable linear combination can be made such that this contribution cancels out.
We will show that this is the case for the sum $G^{\mathrm{th}}_{00}(x_3) + G^{\mathrm{th}}_{11}(x_3)$.
However, this does not necessarily lead to a better agreement between the vacuum and thermal correlation
functions at short distance, due to the presence of constant terms which are not captured in the OPE picture.
This point will be illustrated in the following.

As we discussed in some detail the OPE-based expansion of the electromagnetic-current correlator in a recent publication (\cite{Harris:2020ijy}), 
we largely refer to that for setting the notation and for the introduction of the main observables. 
More specifically, we make reference to the first few paragraphs of Section~3 
for the definition of the Wilson coefficients and of the tower of local twist-two operators, 
and to the beginning of Section~3.1 for the operator mixing in the dimension-four sector.
Moreover, we refer to Appendix~\ref{sect:OPE_test} for the operator expectation values in the theory of free quarks.

\subsection{Leading-order Wilson coefficients}
\label{sub:lo_wilson_coeff}

\begin{figure}[tp]
\includegraphics[width = .85\textwidth]{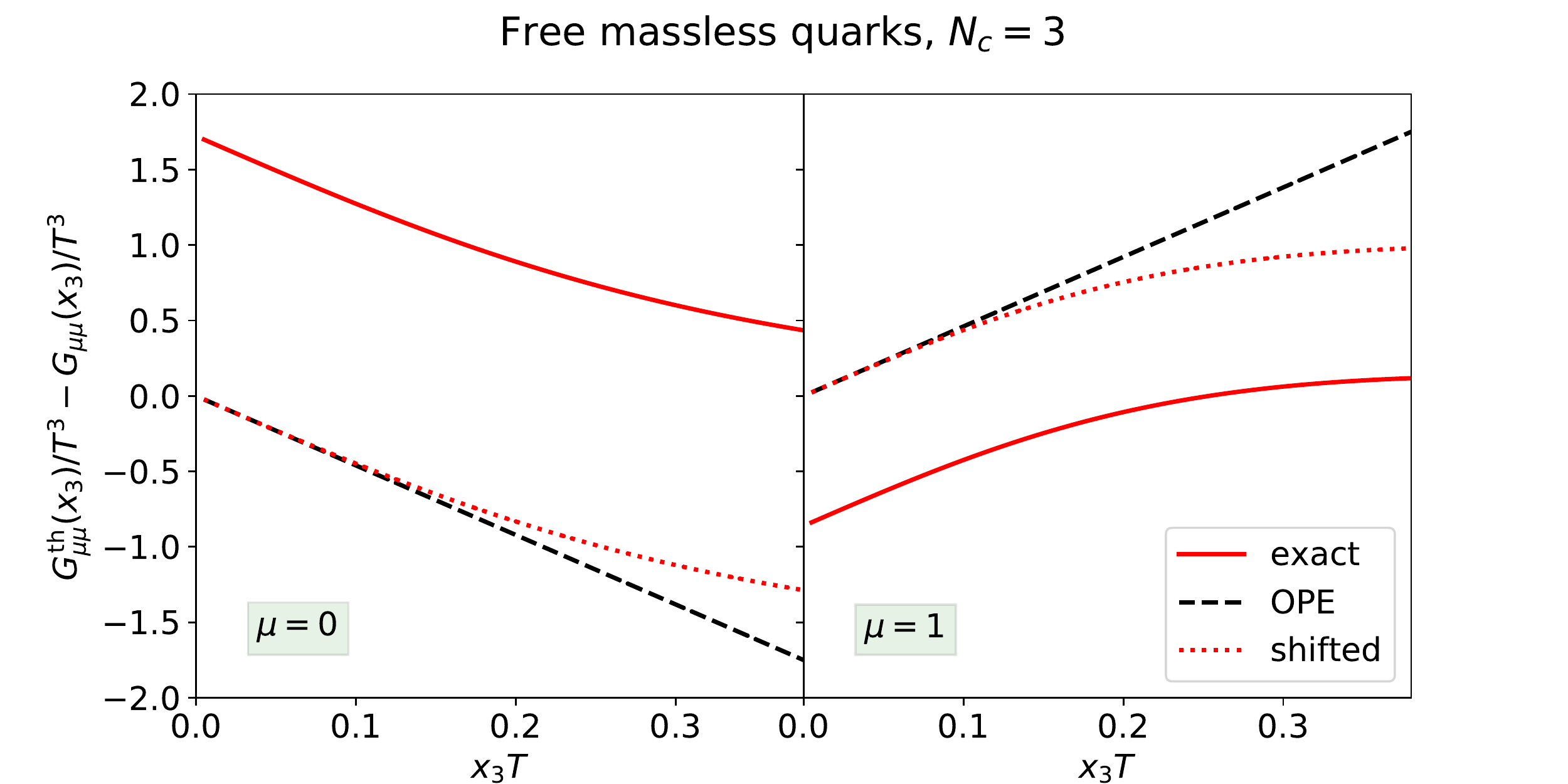}
\caption{Difference between the thermal and vacuum correlation functions 
$G^{\mathrm{th}}_{\mu\mu}(x_3) - G_{\mu\mu}(x_3)$ for $\mu = 0$ (left) and 
$\mu = 1$ (right) at small $x_3$ in the free theory.
The black dashed line represents the OPE prediction for this quantity truncated
at leading order in $x_3$, and the red dotted line is obtained 
by subtracting $[G^{\mathrm{th}}_{\mu\mu} - G_{\mu\mu}]_{x_3=0}$ from the red curve. 
}
\label{fig:constant}
\end{figure}

We start by considering the expansion of the electromagnetic-current correlator in momentum space\footnote{This derivation 
is made with Minkowskian signature. Only in the end we will make the connection to Euclidean correlation functions.}
\begin{equation}
\begin{split}
i \int \dint^4 x \: e^{i q \cdot x}\: \langle {\rm T} \{ J^{\mathrm{em}}_{\mu} (x) J^{\mathrm{em}}_{\nu} (0) \} \rangle
 \overset{\mathrm{LO}}{\sim} \sum_{n=2,4,\dots} \sum_{f} c_{f;\mu \nu \mu_1 \dots \mu_n} (q) \: 
\langle O_{nf}^{\mu_1 \dots \mu_n} \rangle \: .
\end{split}
\label{eq:OPEjj}
\end{equation}
To leading order in the gauge coupling, only the fermionic operators  $O_{nf}^{\mu_1 \dots \mu_n}$
(Eq.~(3.2) in Ref.~\cite{Harris:2020ijy}) contribute, and they are accompanied by the Wilson coefficients
\begin{equation}
\begin{split}
c_{f;\mu \nu \mu_1 \dots \mu_n} (q) = & 
2 Q_f^2 \bigl( -g_{\mu\nu}  + \frac{q_{\mu} q_{\nu}}{q^2} \bigr) 2^n \frac{q_{\mu_1} \dots q_{\mu_n}}{(Q^2)^n} \: + \\
& 2 Q_f^2 \bigl( g_{\mu \mu_1} - \frac{q_{\mu} q_{\mu_1}}{q^2}  \bigr) \bigl( g_{\nu \mu_2} - \frac{q_{\nu} q_{\mu_2}}{q^2}  \bigr)
2^n \frac{q_{\mu_3} \dots q_{\mu_n}}{(Q^2)^{n-1}} 
\: , 
\end{split}
\end{equation}
where $Q_f$ is the electric charge of the quark flavor $f$ and $Q^2 = -q^2$.
The contribution of the dimension-four operator with flavor $f$ is of the form
\begin{equation}
c_{f;\mu\nu\mu_1\mu_2} (q) \langle O_{2f}^{\mu_1\mu_2} \rangle \: ,
\end{equation}
where
\begin{equation}
O_{2f}^{\mu_1\mu_2} = 
\frac{i}{4}\, \Big(\bar\psi_f \gamma^{\{\mu_1}\overleftrightarrow{D}^{\mu_2\}}\psi_f
 - \frac{1}{4}g^{\mu_1\mu_2} \bar\psi_f \overleftrightarrow{D}\!\!\!\!\!/ \;\psi_f\Big)
\: .
\end{equation}
Focusing now on the thermal correlation function,
we express the tensor structure of the operator expectation value as in Ref.~\cite{Harris:2020ijy}
\begin{equation}
\langle O_{2f}^{\mu_1\mu_2} \rangle_T = T^2 \biggl[ u^{\mu_1} u^{\mu_2} - \frac{1}{4} g^{\mu_1\mu_2} \biggr] 
\langle O_{2f} \rangle_T
\: ,
\end{equation}
where $u$ is the four-velocity of the thermal medium.
We fix $q_0 = q_1 = q_2 = 0$, as appropriate for the static screening correlator, and we choose the reference frame in which
the thermal medium is at rest by setting $u = (1,\vec 0)$.
Concentrating on the choices of indices $\mu = \nu = 0,1$, we obtain
\begin{equation}
c_{f;00\mu_1\mu_2} (0,0,0,q_3) \langle O_{2f}^{\mu_1\mu_2} \rangle_T  =
c_{f;11\mu_1\mu_2} (0,0,0,q_3) \langle O_{2f}^{\mu_1\mu_2} \rangle_T  =
\frac{4 Q_f^2 T^2}{q_3^2} \langle O_{2f} \rangle_T
\: .
\label{eq:dim4_LO}
\end{equation}
Fourier-transforming with respect to the variable $q_3$ by using Eq.~(B1) in Ref.~\cite{Chetyrkin:2010dx},
whose infrared regularization results in the $q_3$-independent term being zero,
\begin{equation}
\int_{-\infty}^{\infty} \frac{\dint q_3}{2 \pi}\: \frac{e^{-iq_3 x_3}}{q_3^2} = 
\frac{\Gamma(-1/2)}{\sqrt{4\pi} \: \Gamma(1)} \biggl( \frac{x_3^2}{4} \biggr)^{1/2} = - \frac{\abs{x_3}}{2}
\: ,
\label{eq:FT_1overq2}
\end{equation}
we find for the coordinate-space version of Eq.\ (\ref{eq:dim4_LO}) 
\begin{equation}
\sum_f c_{f;00\mu_1\mu_2} (x_3) \langle O_{2f}^{\mu_1\mu_2} \rangle_T = 
\sum_f c_{f;11\mu_1\mu_2} (x_3) \langle O_{2f}^{\mu_1\mu_2} \rangle_T =
-2 \abs{x_3} \bigl( \sum_f Q_f^2 \bigr) T^2 \langle O_{2f} \rangle_T
\: .
\end{equation}
Finally, to get to the isovector vector-current correlators in Euclidean space-time we drop the factor $\sum_f Q_f^2$ and
multiply by $-1$ the contribution to the correlator with two spatial indices. 
We obtain the OPE prediction to leading order in the gauge coupling
\begin{equation}
G^{\mathrm{th}}_{00} (x_3) \overset{x_3 \to 0}{\sim} G_{00} (x_3) -2 \abs{x_3} T^2 \langle O_{2f} \rangle_T + O(\abs{x_3}^3)
\label{eq:OPE00}
\end{equation}
\begin{equation}
G^{\mathrm{th}}_{11} (x_3) \overset{x_3 \to 0}{\sim} G_{11} (x_3) +2 \abs{x_3} T^2 \langle O_{2f} \rangle_T + O(\abs{x_3}^3)
\: .
\label{eq:OPE11}
\end{equation}
In Appendix~\ref{sect:OPE_test} we verify the correctness of the $O(\abs{x_3})$ term in the theory of free quarks.
We observe that the term linear in $\abs{x_3}$ cancels out in the sum $G^{\mathrm{th}}_{00}(x_3) + G^{\mathrm{th}}_{11}(x_3)$.
This fact is not sufficient to conclude that taking this combination improves the short-distance agreement with 
the vacuum correlation function $G_{00}(x_3) = G_{11}(x_3)$, 
because the OPE is not able to capture constant terms in the small-$x_3$ expansion, 
as those are not of short-distance origin.
The free-theory correlators  provide a concrete example of this situation,
as shown in Figure~\ref{fig:constant}.
The figure shows the difference between the thermal correlation functions
$G^{\mathrm{th}}_{00}(x_3)$, $G^{\mathrm{th}}_{11}(x_3)$ and their vacuum counterparts
as a function of $x_3$, between $x_3 = 0$ and $x_3 \simeq 0.4/T$.
The leading OPE term, as explicitly computed in Appendix~\ref{sect:OPE_test},
is also reported.
Contrary to the OPE prediction, the difference between the thermal and vacuum correlators
starts from a nonzero value in $x_3 = 0$. However, once we subtract $\Delta G_0 \equiv [G^{\mathrm{th}} - G]_{x_3 = 0}$,
we find that the truncated OPE provides indeed a good description of 
$(G^{\mathrm{th}} - G - \Delta G_0)$ at small $x_3$.

\subsection{Next-to-leading-order Wilson coefficients}
\label{sub:nlo_wilson_coeff}

To higher order in perturbation theory, the mixing under renormalization with the gluonic
twist-two operators $O_{ng}^{\mu_1 \dots \mu_n}$ (Eq.~(3.3) in Ref.~\cite{Harris:2020ijy}) must be taken into account.
To write explicitly the contribution from the dimension-four operators to NLO precision, 
we follow closely the sections~3 and~3.1 of Ref.~\cite{Harris:2020ijy}
and refer to those for any unexplained notation.
Going back to Eq.~(\ref{eq:dim4_LO}), its NLO equivalent is obtained by making the substitution
\begin{equation}
\begin{split}
\langle O_{2f} \rangle_T \: \longrightarrow \: 
& \frac{1}{2} \frac{1}{16/3 + N_f} \frac{e+p}{T^2} + \\
& \frac{1}{N_f(16/3 + N_f)} \biggl( \frac{\log(q_3^2/\Lambda^2)}{\log(\tilde \mu^2/\Lambda^2)} 
\biggr)^{\gamma} \biggl( \frac{16}{3} \sum_{f'}
\langle O_{2f'} \rangle_T - N_f \langle O_{2g} \rangle_T \biggl)
\: ,
\end{split}
\end{equation}
where $e$ and $p$ are the energy density and the pressure of the thermal
medium and $\tilde \mu$ and $\Lambda$ represent two energy scales, respectively the one
at which the local operators are renormalized and the one at which the one-loop renormalized 
coupling diverges.
The power $\gamma$ is one of the eigenvalues of the anomalous-dimension matrix,
\begin{equation}
\gamma = -\frac{4}{3} \biggl( \frac{16}{3} + N_f \biggr) \frac{1}{2b_0} \overset{N_f = 2}{=} -0.51
\: ,
\end{equation}
where $b_0 = 11 - \frac{2}{3} N_f$ is the coefficient of the one-loop contribution to
the beta function. 

To get the final expression in position space, we are faced with the problem of computing
the Fourier transform 
\begin{equation}
\frac{1}{2\pi} \int_{-\infty}^{\infty} \dint q_3 \: e^{-i q_3 x_3} \: \frac{\log(q_3^2/\Lambda^2)^{\gamma}}{q_3^2}
\: .
\label{eq:ft_log}
\end{equation}
As in the case of Eq.~(\ref{eq:FT_1overq2}), the Fourier integral is divergent and
a regularization procedure is needed to correctly extract the asymptotic dependence on $x_3$.
We discuss this point in detail in Appendix~\ref{app:FT_log} and report here the final result
\begin{equation}
\frac{\log(q_3^2/\Lambda^2)^{\gamma}}{q_3^2} \: \overset{\mathrm{coord. \:\: space}}{\longrightarrow} \:
-\frac{\abs{x_3}}{2} \log(1/(x_3\Lambda)^2)^{\gamma}
\: .
\end{equation}
The OPE prediction to NLO in the gauge coupling reads
\begin{equation}
\begin{split}
G^{\mathrm{th}}_{\mu\mu} (x_3) - G_{\mu\mu} (x_3) \overset{\mathrm{NLO}}{\sim}
& s_{\mu} 2 \abs{x_3} T^2 \biggl\{ \frac{1}{2} \frac{1}{16/3 + N_f} \frac{e+p}{T^2} + \\
& \frac{1}{N_f(16/3 + N_f)} \biggl( \frac{\log(1/(x_3\Lambda)^2)}{\log(\tilde \mu^2/\Lambda^2)} 
\biggr)^{\gamma} \biggl( \frac{16}{3} \sum_{f'}
\langle O_{2f'} \rangle_T - N_f \langle O_{2g} \rangle_T \biggl) \biggr\} + O(\abs{x_3}^3)
\: ,
\end{split}
\end{equation}
where $\mu \neq 3$ and $s_{\mu}$ is a sign factor evaluating to 1 for $\mu = 1,2$ and to $-1$ for $\mu = 0$.
While we did not discuss explicitly the case $\mu = 2$ so far, we point out that the  
symmetries of the screening correlator constrain it to be equal to the case $\mu = 1$.
In the extreme $x_3 \to 0$ limit the logarithmic contribution is subleading, and we find
\begin{equation}
\bigl[ G^{\mathrm{th}}_{\mu\mu} (x_3) - G_{\mu\mu} (x_3) \bigr]_{\mathrm{NLO}} \overset{x_3 \to 0}{\sim}
s_{\mu} \abs{x_3} \frac{e+p}{16/3 + N_f}  \: , \quad \mu \neq 3
\: .
\end{equation}
As already pointed out in Ref.~\cite{Harris:2020ijy} relative to the structure functions of the quark-gluon plasma,
the free-theory prediction of Eqs.~(\ref{eq:OPE00}), (\ref{eq:OPE11}) is not recovered in the short-distance limit $x_3 \to 0$, 
contrary to the intuition coming from the property of asymptotic freedom.
As a consequence of the mixing between operators under renormalization, the free theory represents here an extreme case which
is not connected to the real interacting theory by an expansion in powers of the QCD coupling.

\section{Test of the leading OPE prediction in the free theory}
\label{sect:OPE_test}

The expressions (\ref{eq:OPE00}), (\ref{eq:OPE11}) can be verified in the theory of free quarks.
In this case, the expectation value of the dimension-four operator reads (see Appendix~B of Ref.~\cite{Harris:2020ijy})
\begin{equation}
\langle O_{2f} \rangle^{\mathrm{free}}_T = \frac{7 \pi^2 T^2 N_c}{90}
\: ,
\label{eq:O2f_free}
\end{equation}
where $N_c$ is the number of colors.
We can verify the OPE prediction by expanding the free correlation function in powers of $\abs{x_3}$.
To do this, we make use of Eqs.~(3.5) and (3.6) in Ref.~\cite{Brandt:2014uda}, which give the free thermal vector
correlator projected to Matsubara frequency $k_n$ as a function of the spatial coordinates $\vec r = (x_1,x_2,x_3)$.
After identifying the relevant terms in the expansion in powers of $\vec r$, 
we project to zero momentum in the directions $x_1$ and $x_2$ to obtain the
corresponding contribution to the screening correlator.
In the static sector $k_n = 0$, we have
\begin{equation}
G_{00}^{(k_n = 0)} (\vec r) = -\frac{N_c T^3}{r^2} \biggl( \frac{\cosh \bar r}{\bar r \sinh^2 \bar r} +
\frac{1}{{\bar r}^2 \sinh \bar r} \biggr)
\: ,
\label{eq:G00free_coordsp}
\end{equation}
and
\begin{equation}
\begin{split}
G_{11}^{(k_n = 0)} (\vec r) = 
\frac{N_c T^3}{r^2} \biggr\{ & \frac{x_1^2}{r^2} \biggr[ \frac{\cosh \bar r}{\bar r \sinh^2 \bar r}
+ \frac{1}{{\bar r}^2 \sinh \bar r} \biggr] \\
& - \biggl( 1 - \frac{x_1^2}{r^2} \biggr) \biggl[ \frac{\cosh \bar r}{\bar r \sinh^2 \bar r} + \frac{1}{{\bar r}^2 \sinh \bar r}
+ \frac{1}{2 \sinh \bar r} + \frac{1}{\sinh^3 \bar r} \biggr] \biggr\}
\: ,
\end{split}
\label{eq:G11free_coordsp}
\end{equation}
where $\bar r \equiv 2 \pi T r$ and $r \equiv \abs{\vec r}$.
We start by expanding Eq.~(\ref{eq:G00free_coordsp}) in powers of the spatial coordinates.
In the OPE picture, we expect a contribution from the dimension-four operator of the form
\begin{equation}
\frac{\tilde c_{\mu_1\mu_2} \langle O_{2f}^{\mu_1\mu_2} \rangle}{r}
\: ,
\end{equation}
where $\tilde c_{\mu_1\mu_2}$ is a dimensionless coefficient.
Therefore we are interested in the term proportional to $1/r$ in the expansion
\begin{equation}
G_{00}^{(k_n = 0)} (\vec r) = -\frac{N_c T^4}{r} \: \frac{2 \pi}{{\bar r}^4} \biggl(
2 - \frac{7}{180} {\bar r}^4 + O({\bar r}^6) \biggr)
\: ,
\end{equation}
which is 
\begin{equation}
\frac{7 \pi N_c T^4}{90} \frac{1}{r} \: .
\end{equation}
To project to zero momentum in the directions $x_1$ and $x_2$, we make use of the following equation
\begin{equation}
\begin{split}
\int_{-\infty}^{\infty} \frac{\dint x_1 \dint x_2}{r} & = 
2 \pi \lim_{L \to \infty} \int_0^{L} \frac{\rho \: \dint \rho}{\sqrt{\rho^2+x_3^2}} =
2 \pi \lim_{L \to \infty} \bigl( - \abs{x_3} + \sqrt{L^2 + x_3^2} \bigr) \\
& = -2 \pi \abs{x_3} + \mathrm{proportional \: to \: spatial \: extent,\: independent \: of \:}x_3
\: .
\end{split}
\end{equation}
To conclude, the linear term in the expansion of the correlator $G^{\mathrm{th}}_{00}(x_3)$ in powers of $\abs{x_3}$ is given by
\begin{equation}
-\frac{7 \pi^2 N_c T^4}{45} \abs{x_3}
\: ,
\end{equation}
and it is in agreement with the OPE prediction $-2 \abs{x_3} T^2 \langle O_{2f} \rangle_T^{\mathrm{free}}$.

We now repeat the same procedure for the correlator~(\ref{eq:G11free_coordsp}).
In this case, the contribution from the dimension-four operator is of the form
\begin{equation}
\biggl( \frac{\tilde c_{\mu_1\mu_2}}{r} + \frac{x_1^2}{r^3} \tilde d_{\mu_1\mu_2} \biggr)
\langle O_{2f}^{\mu_1\mu_2} \rangle 
\: ,
\end{equation}
where $\tilde c_{\mu_1\mu_2}$ and $\tilde d_{\mu_1\mu_2}$ are dimensionless coefficients.
From the expansion
\begin{equation}
\begin{split}
G_{11}^{(k_n = 0)} (\vec r) = 
N_c T^3 \biggr\{ & \frac{x_1^2}{r^3} \: \frac{2 \pi T}{{\bar r}^4} \biggr[ 2 - \frac{7}{180} {\bar r}^4  +
O({\bar r}^6) \biggr] \\
& - \biggl( \frac{1}{r} - \frac{x_1^2}{r^3} \biggr) \frac{2 \pi T}{{\bar r}^4}
\biggl[ 3 + \frac{7}{360} {\bar r}^4 + O({\bar r}^6)
\biggr] \biggr\}
\: ,
\end{split}
\end{equation}
we see that the terms proportional to $1/r$ and $x_1^2/r^3$ are given by
\begin{equation}
- \frac{7 \pi N_c T^4}{180} \biggl( \frac{1}{r} + \frac{x_1^2}{r^3} \biggr)
\: .
\end{equation}
We project to zero momentum in the directions $x_1$ and $x_2$ by applying the following procedure
\begin{equation}
\begin{split}
\int_{-\infty}^{\infty} \dint x_1 \dint x_2 \: \biggl( \frac{1}{r} + \frac{x_1^2}{r^3} \biggr) & =
2 \pi \lim_{L \to \infty} \biggl( \int_0^{L} \dint \rho \frac{\rho}{\sqrt{\rho^2 + x_3^2}}
+ \frac{1}{2} \int_0^{L} \dint \rho \frac{\rho^3}{(\rho^2 + x_3^2)^{\frac{3}{2}}} \biggr) \\
& = 2 \pi \lim_{L \to \infty} \biggl( - \abs{x_3} + \sqrt{L^2 + x_3^2} + 
\frac{1}{2} \frac{L^2 + 2 x_3^2 - 2 \abs{x_3} \sqrt{L^2 + x_3^2}}{\sqrt{L^2 + x_3^2}} \biggr) \\
& = - 4 \pi \abs{x_3} + \mathrm{proportional \: to \: spatial \: extent, \: independent \: of \:} x_3
\: .
\end{split}
\end{equation}
To conclude, the linear term in the expansion of $G^{\mathrm{th}}_{11}(x_3)$ in powers of $\abs{x_3}$ is
\begin{equation}
\frac{7 \pi^2 N_c T^4}{45} \abs{x_3}
\end{equation}
and it agrees with the OPE prediction $2 \abs{x_3} T^2 \langle O_{2f} \rangle_T^{\mathrm{free}}$.

\section{Fourier transform of \texorpdfstring{$\log(q^2/\Lambda^2)^{\gamma}/q^2$}{log}}
\label{app:FT_log}

The authors of Ref.~\cite{Wong:1978} study the the asymptotic behaviour for $q\to\infty$ of the one-sided Fourier transform
\begin{equation}
  F(q) = \int_0^\infty \dd{x} e^{iqx} f(x)
\end{equation}
of a function that has logarithmic singularities for $x\to 0^+$
\begin{equation}
  f(x) \sim \sum_{m=0}^{\infty} c_m x^{\alpha_m-1}(-\log x)^{\beta_m} ,
\end{equation}
where $\alpha_m\to +\infty$ as $m\to\infty$ with $\Re\alpha_{m+1}\geq\Re\alpha_m$, and the $\beta_m$ are arbitrary complex numbers.
The main result is that for $q\to\infty$ holds
\begin{equation}
  F(q) = \sum_{m=0}^{M-1} c_m J(\alpha_m, \beta_m, q) + o(q^{-n})
\end{equation}
where $M$ is a positive integer such that $\Re\alpha_{M-1}\leq n<\Re\alpha_M$ and
\begin{equation}
  J(\alpha, \beta, q) = \int_0^{\infty e^{i\theta}} \dd{x} e^{iqx} x^{\alpha-1} (-\log x)^\beta \sim \frac{e^{\alpha\pi i/2}}{q^\alpha} \sum_{r=0}^\infty c_r(\alpha, \beta) (\log q)^{\beta-r} ,
\end{equation}
with the $c_r(\alpha, \beta)$ given by
\begin{equation}
   c_r(\alpha, \beta) = (-1)^r {\beta\choose r} \sum_{k=0}^r {r\choose k} \Gamma^{(k)}(\alpha) \left(\frac{\pi i}{2}\right)^{r-k} .
\end{equation}
Setting $M=1$, $\alpha_0=2$ and $\beta_0=\gamma$, and using $c_0(2, \gamma)=1$ we obtain
\begin{equation}
  f(x) \stackrel{x\to 0^+}{\sim} x(-\log x)^\gamma \quad \Rightarrow \quad F(q) \stackrel{q\to\infty}{\sim} -\frac{(\log q)^\gamma}{q^2} \left[ 1 + \sum_{r=1}^\infty c_r(2, \gamma) (\log q)^{-r} \right] + o(q^{-2}) .
\end{equation}
An arbitrary number of subleading $(\log q)^r$ terms in the asymptotic behaviour of $F(q)$ can be removed with appropriately chosen $\alpha_m$ and $\beta_m$.
In particular, we found that setting $\alpha_m=\alpha$, $\beta_m=\gamma-m$ and
\begin{equation}
  d_m(\alpha, \gamma) = (-1)^r {\gamma\choose m} \sum_{k=0}^m {m\choose k} \Gamma(\alpha) \dv[k]{}{\alpha} \frac{1}{\Gamma(\alpha)} \left(\frac{\pi}{2i}\right)^{m-k}
\end{equation}
we obtain
\begin{equation}
  f(x) \stackrel{x\to 0^+}{\sim} \sum_{m=0}^{\infty} d_m(\alpha, \gamma) x^{\alpha-1}(-\log x)^{\gamma-m} \quad \Rightarrow \quad F(q) \stackrel{q\to\infty}{\sim} \frac{e^{\alpha\pi i/2}}{q^\alpha} (\log q)^\gamma + o(q^{-\alpha}) ,
\end{equation}
or, in the $\alpha=2$ case,
\begin{equation}
  f(x) \stackrel{x\to 0^+}{\sim} \sum_{m=0}^{\infty} d_m(2, \gamma) x(-\log x)^{\gamma-m} \quad \Rightarrow \quad F(q) \stackrel{q\to\infty}{\sim} -\frac{(\log q)^\gamma}{q^2} + o(q^{-2}) .
\end{equation}
From this it is easy to show that for $\gamma\in\mathbb{R}$ a function with the asymptotic behaviour
\begin{equation}
  \frac{(\log q^2/\Lambda^2)^\gamma}{q^2} \qquad \text{for $q\to \infty$}
\end{equation}
is the two-sided Fourier transform of a function that goes like
\begin{equation}
  -\frac{\abs{x}}{2} \left(\log\frac{1}{(\abs{x}\Lambda)^2}\right)^\gamma \left[ 1 + \sum_{m=1}^{\infty} \Re d_m(2, \gamma) \left(\frac{1}{2}\log\frac{1}{(\abs{x}\Lambda)^2}\right)^{-m} \right] \qquad \text{for $x\to 0$}.
\end{equation}
The first coefficients in the sum in the r.h.s.\ evaluates to
\begin{gather}
  \Re d_1(2, \gamma) = 2(1-\gamma_E)\gamma \approx \num{0.422784}\gamma , \\
  \Re d_2(2, \gamma) \approx \num{-1.46679}\gamma(\gamma-1) , \\
  \Re d_3(2, \gamma) \approx \num{-0.71268}\gamma(\gamma-1)(\gamma-2) , \\
  \Re d_3(2, \gamma) \approx \num{0.516685}\gamma(\gamma-1)(\gamma-2)(\gamma-3) .
\end{gather}

\section{Details on the lattice free-theory computation}
\label{app:free_corr}

In this appendix we collect some details on the free-theory computation.
In the theory of non-interacting massless Wilson quarks, defined on a
spatially-infinite lattice, the zero-temperature quark propagator can be expressed as
\begin{equation}
\begin{split}
\langle \psi_f (x) \bar\psi_f(y) \rangle =  
& \int_{-\frac{\pi}{a}}^{\frac{\pi}{a}} \frac{\dint^3 p}{(2 \pi)^3} \:
\frac{e^{-\omega_{\vec p} \abs{x_3-y_3} + i \vec p \cdot (\vec x - \vec y)}}{D(\vec p)} \times \\
& \biggl( \sgn (x_3 - y_3) \frac{1}{a} \sinh(a \omega_{\vec p}) \g_3 - i \vec \g \cdot \mathring{\vec p} + C(\vec p) +
\delta_{x_3,y_3} \frac{1}{a} \sinh (a \omega_{\vec p}) \biggr) \: ,
\end{split}
\label{eq:prop}
\end{equation}
where
\begin{equation}
\mathring{p}_{\mu} = \frac{1}{a} \sin (a p_{\mu}) \: , \quad
\hat p_{\mu} = \frac{2}{a} \sin \bigl( \frac{a p_{\mu}}{2} \bigr) \: , 
\end{equation}
\begin{equation}
A(\vec p) = 1 + \frac{1}{2} a^2 \hat{\vec p}^{\: 2} \: , 
\quad
B(\vec p) = \hat{\vec p}^{\: 2} + \frac{1}{2} a^2 \sum_{k<l} {\hat p}^2_k {\hat p}^2_l \: ,
\quad
C(\vec p) = \frac{a}{2} \biggl( \hat{\vec p}^{\: 2} - \frac{B(\vec p)}{A(\vec p)} \biggr) \: ,
\end{equation}
\begin{equation}
D(\vec p) = \sqrt{B(\vec p) \bigl(4 A(\vec p) + a^2 B(\vec p) \bigr) } = \frac{2}{a} A(\vec p) \sinh(a \omega_{\vec p}) 
\: .
\end{equation}
The sign function in Eq.~(\ref{eq:prop}) evaluates to zero when its argument is zero, 
and the vector $\vec p$ has three components denoted by $(p_0, p_1, p_2)$.
Similarly, $\vec \gamma = (\gamma_0,\gamma_1,\gamma_2)$ and $\vec x = (x_0,x_1,x_2)$.
At finite temperature $T$, the thermal propagator 
$\langle \psi_f(x) \bar\psi_f(y) \rangle_T$ can be obtained from Eq.~(\ref{eq:prop}) by 
assigning to $p_0$ discrete values corresponding to the fermionic Matsubara frequencies $p_0 = (2 n + 1)\pi T$,
and by making the substitution
\begin{equation}
\int_{-\frac{\pi}{a}}^{\frac{\pi}{a}} \frac{\dint p_0}{2 \pi} \to T \sum_{n=0}^{N_t-1} 
\: ,
\label{eq:p0toMats}
\end{equation}
where $N_t = 1/(aT)$ is the number of lattice points in the Euclidean-time direction.
With the propagators at hand, we have all the elements to compute the correlation functions
(\ref{eq:gvac}) and (\ref{eq:gth}) in the massless free theory.
In the vacuum, and for $\mu = \nu \neq 3$, we have
\begin{equation}
\begin{split}
\mathcal G_{\mu \mu} (x_3) \overset{\mu \neq 3}{=} 4 N_c \int_{-\frac{\pi}{a}}^{\frac{\pi}{a}} 
\frac{\dint^3 p}{(2 \pi)^3} \frac{e^{-2 \omega_{\vec p} \abs{x_3}}}{D(\vec p)^2}
& \biggl[ 2 \mathring p_{\mu} \sin(a p_{\mu}) \bigl( C(\vec p) + \delta_{x_3,0} \frac{1}{a} \sinh(a \omega_{\vec p}) \bigr) \\
& - \cos(a p_{\mu}) \biggl( \sgn(x_3)^2 \frac{1}{a^2} \sinh^2(a \omega_{\vec p}) - 2 {\mathring p_{\mu}}^2
+ \mathring{\vec p}^2 \\
& + \bigl( C(\vec p) + \delta_{x_3,0} \frac{1}{a} \sinh(a \omega_{\vec p}) \bigr)^2 \biggr) \biggr] \: ,
\end{split}
\label{eq:Gvac_free}
\end{equation}
where $N_c$ represents the number of colors. 
In the vacuum and in infinite volume there is no difference between
projecting to zero momentum in the directions $(x_1,x_2,x_3)$ or $(x_0,x_1,x_2)$.
In this appendix we choose the second option, 
in order to keep a closer analogy with the thermal screening correlator.
The thermal correlation function $\mathcal G^{\mathrm{th}}_{\mu\mu}(x_3)$,
with $\mu \neq 3$, can be obtained starting from Eq.~(\ref{eq:Gvac_free})
and exchanging the integral over $p_0$ with a sum over fermionic
Matsubara modes, as in Eq.~(\ref{eq:p0toMats}).

We consider the fourth-moment observable $\mathcal I (t)$
 as defined in Eq.~(\ref{eq:x04moment_lat}).
In the limit $a \to 0$ its integrand reads
\begin{equation}
\begin{split}
x_3^4 \: \mathcal G(x_3) = 
 x_3^4 \int_{-\frac{\pi}{a}}^{\frac{\pi}{a}} \frac{\dint^3 p}{(2 \pi)^3} 
\: e^{-2 p \abs{x_3}} \biggl[ \hat f_{0,0} (\hat{\vec p})\: + 
 a^2 \bigl( p^2 \hat f_{2,0}(\hat{\vec p}) + 
\abs{x_3} p^3 \hat f_{2,1}(\hat{\vec p}) \bigr) + O(a^4) \biggr] \: ,
\end{split}
\label{eq:expansion}
\end{equation}
where
\begin{equation}
\hat f_{0,0} (\hat{\vec p}) = 2 N_c \biggl( 1 - \frac{p_1^2}{p^2} \biggr)
\: ,
\end{equation}
\begin{equation}
\hat f_{2,0} (\hat{\vec p}) = - 2 N_c \biggl[ 1 - \frac{1}{2} \frac{p_1^2}{p^2} \biggl( 1 -
\frac{2}{3} \: \frac{p_0^4 + p_1^4 + p_2^4}{p^4} + \frac{5}{3} \: \frac{p_1^2}{p^2} \biggr)  \biggr]
\: ,
\end{equation}
\begin{equation}
\hat f_{2,1} (\hat{\vec p}) = N_c \: \frac{2}{3} \biggl( 1 - \frac{p_1^2}{p^2} \biggr) \biggl( 1 + \frac{p_0^4 + p_1^4 + p_2^4}{p^4} \biggr)
\: .
\end{equation}
As before we have $\vec p = (p_0,p_1,p_2)$, and we  introduced the 
notation $p \equiv \abs{\vec p}$.
We observe that a generic term of the expansion within square brackets in 
Eq.~(\ref{eq:expansion}) can be expressed as
\begin{equation}
a^n \abs{x_3}^m p^{n+m} \hat f_{n,m}(\hat{\vec p}) \: ,
\label{eq:expansion_general}
\end{equation}
with $n = 2,4,\dots$, $m \geq 0$ and where the dimensionless function $\hat f_{n,m}(\hat{\vec p})$ 
contains the dependence on the orientation of the vector $\vec p$ ($\hat{\vec p} \equiv \vec p/p$).

\section{Analysis of \texorpdfstring{$\Delta_2 \Pi(Q^2)$}{D2Pi} in the theory of free quarks \la{sec:Delta2Pi}}

As an outlook towards future applications, a strategy to compute the hadronic contribution
to the running of the electromagnetic coupling up to the $Z$ mass is outlined in Section~\ref{sec:concl}.
In this appendix, we test the core of this strategy in the theory of massless free quarks,
whose lattice formulation is described in detail in Section~\ref{sect:latt_form} and in Appendix~\ref{app:free_corr}.
The observable under analysis is
\begin{equation}
\begin{split}
\Delta_2 \Pi (Q^2) & \equiv \Pi(Q^2) - \Pi(Q^2/4) \\
& = \int_0^{\infty} \dint x_3 G(x_3) \frac{4}{Q^2} \biggl[ 4 \sin^2 \biggl(\frac{Q x_3}{4} \biggr)
- \sin^2 \biggl(\frac{Q x_3}{2} \biggl) \biggl]
\: ,
\end{split}
\label{eq:deltaPi_def}
\end{equation}
where $\Pi(Q^2)$ is the hadronic vacuum polarization and the correlation function $G(x_3)$ is defined in Eq.~(\ref{eq:G}).
The thermal equivalents at two different temperatures $\Delta_2^{\mathrm{th}} \Pi(Q^2;T)$, 
$\Delta_2^{\mathrm{th}} \Pi(Q^2;T/2)$ are also considered.
The temperature is fixed to $T = Q/(8 \pi)$.
In view of computing this observable on the lattice, an upper cut $x_3^{\mathrm{cut}} = 2/T$ is set in the integral
of Eq.~(\ref{eq:deltaPi_def}). The lattice observables are denoted by $\Delta_2\mathbf{\Pi} (Q^2)$, 
$\Delta_2^{\mathrm{th}}\mathbf{\Pi} (Q^2;T)$.
As in the analysis presented in Section~\ref{sec:pert}, we fix the physical value of the temperature to
$T = 246.25$~MeV, which is close to the temperature of the QCD ensembles considered in this study,
and which assigns to the lattice Euclidean-time direction the physical extent $1/T = 0.8$~fm.
As a consequence, $Q = 8 \pi T \simeq 6.2$~GeV.

\begin{figure}[tp]
\includegraphics[width = .5\textwidth]{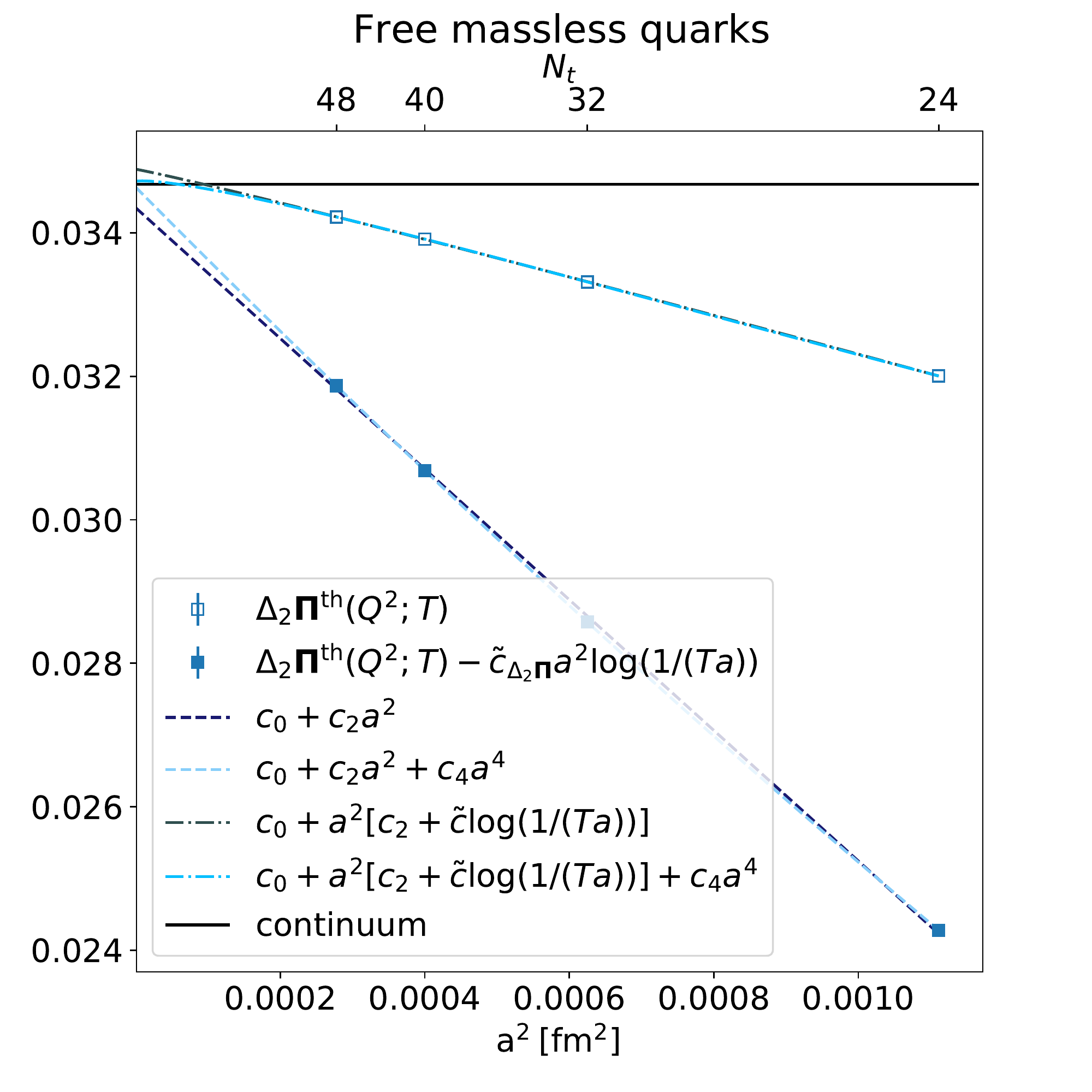}~
\includegraphics[width = .5\textwidth]{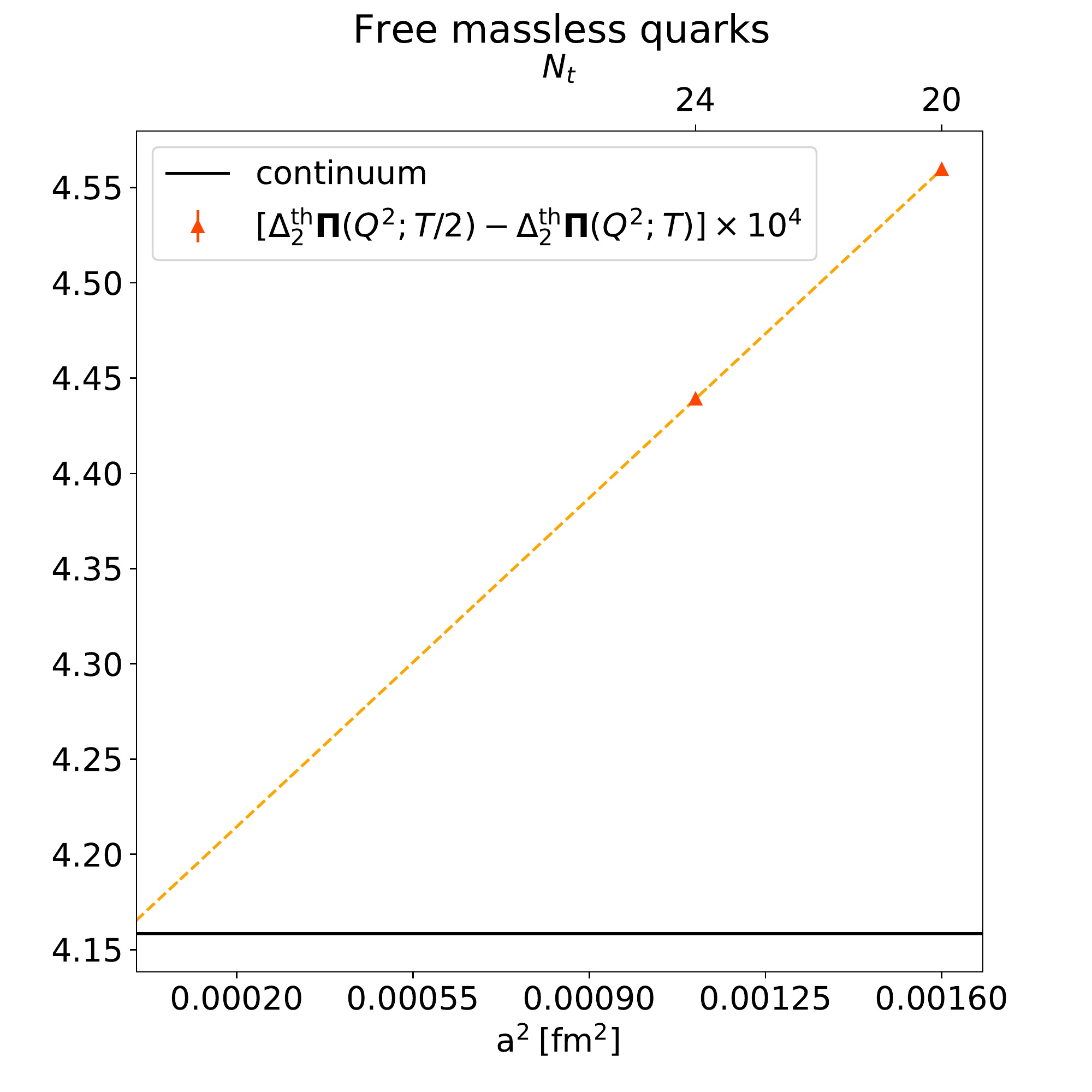}
\caption{Left: continuum extrapolation of the thermal observable $\Delta_2^{\mathrm{th}}\mathbf{\Pi}(Q^2;T)$,
with $T = 246.25$~MeV and  $Q = 8\pi T \simeq 6.2$~GeV.
The $O(a^2 \log(1/a))$ lattice artifact is either included in the fit ansatz or explicitly subtracted by using 
the known prefactor $\tilde c_{\Delta_2\mathbf{\Pi}}$ (\ref{eq:clog_DeltaPi}).
The accuracy of the resulting continuum estimates is reported in Table~\ref{tab:deltaPi}. 
Right: continuum extrapolation of the bias correction
$[\Delta_2^{\mathrm{th}} \mathbf{\Pi} (Q^2;T/2) - \Delta_2^{\mathrm{th}} \mathbf{\Pi}(Q^2;T)]$,
together with its known continuum value. The fit ansatz is linear in $a^2$.
}
\label{fig:deltaPi}
\end{figure}

The first step consists in obtaining a continuum estimate of the thermal observable 
$\Delta_2 \mathbf \Pi (Q^2;T)$. With this goal, we consider four lattices with temperature $T$ and with $N_t = 24,32,40,48$
lattice sites in the Euclidean-time direction. The temperature and the lattice spacing are related by
$T = 1/(a N_t)$. 
With this setup, and making good use of the knowledge about the logarithmic lattice artifacts (see Section~\ref{sect:log}),
we obtain a continuum estimate differing from the correct continuum value by only $0.1\%$.
The lattice data and the fit curves are shown in the left panel of Figure~\ref{fig:deltaPi} together with the
correct continuum value, the accuracy of the resulting continuum estimates is reported in Table~\ref{tab:deltaPi}.
The prefactor of the $O(a^2\log(1/a))$ cutoff effect, which we denote by $\tilde c_{\Delta_2\mathbf{\Pi}}$,
can be computed by observing that the short-distance limit of the integration kernel is
\begin{equation}
\frac{4}{Q^2} \biggl[ 4 \sin^2 \biggl(\frac{Q x_3}{4} \biggr)
- \sin^2 \biggl(\frac{Q x_3}{2} \biggl) \biggl]
\overset{x_3 \to 0}{\sim} \frac{Q^2}{16} x_3^4
\: ,
\end{equation}
which implies
\begin{equation}
\tilde c_{\Delta_2\mathbf{\Pi}} = \frac{Q^2}{16} \tilde c_{\mathcal I} = \frac{7 Q^2}{320 \pi^2}
\: .
\label{eq:clog_DeltaPi}
\end{equation}
The value of $\tilde c_{\mathcal I}$ is given in Eq.~(\ref{eq:clog}). 

\begin{table}[tp]
\center
\caption{
Accuracy of the continuum extrapolation of the thermal quantity $\Delta_2^{\mathrm{th}}\mathbf{\Pi}(Q^2;T)$,
measured against the correct continuum value.
The label ``plain'' refers to fitting the plain lattice observable, while for the case ``subtr.'' the
logarithmic lattice artifact is subtracted as $\Delta_2^{\mathrm{th}}\mathbf{\Pi}(Q^2;T) - \tilde c_{\Delta_2\mathbf{\Pi}} a^2\log(1/(Ta))$. 
For the choice of scales made here $T = 246.25$~MeV, $Q = 8\pi T \simeq 6.2$~GeV, we have 
$\tilde c_{\Delta_2\mathbf{\Pi}} \simeq 2.2~\mathrm{fm}^{-2}$ (see Eq.~(\ref{eq:clog_DeltaPi})). 
The lattice data and the fit curves are shown in Figure~\ref{fig:deltaPi}, left panel.
}
\begin{tabular}{c c c c c}
\toprule
\multicolumn{2}{c}{$\abs{c_0 - \Delta_2^{\mathrm{th}}\Pi}/\Delta_2^{\mathrm{th}}\Pi\:\:\:$} & $\tilde c~[\mathrm{fm}^{-2}]$ & ansatz  \\
\midrule
 0.8\%   & 0.9\%    & --   & $c_0 + c_2 a^2$                                       \\
 0.7\%   & 0.1 \%   & --   & $c_0 + c_2 a^2 + c_4 a^4$                             \\
 0.6\%   & --       & 0.29 & $c_0 + a^2 [c_2 + \tilde c \log(1/(Ta))]$             \\
 0.1\%   & --       & 1.6  & $c_0 + a^2 [c_2 + \tilde c \log(1/(Ta))] + c_4 a^4$   \\
\midrule
plain & subtr. &  &  \\
\bottomrule
\end{tabular}
\label{tab:deltaPi}
\end{table}

The difference between the thermal observable 
$\Delta_2^{\mathrm{th}}\Pi(Q^2;T)$ and its vacuum counterpart is of about one percent.
To correct for this bias, we add a continuum estimate of the difference
$\Delta_2^{\mathrm{th}}\mathbf{\Pi}(Q^2;T/2) - \Delta_2^{\mathrm{th}}\mathbf{\Pi}(Q^2;T)$.
Within the desired precision, this is a good-enough approximation
of the difference between the vacuum observable and $\Delta_2^{\mathrm{th}}\Pi(Q^2;T)$, and it has
the advantage of being cheaper to compute at equal lattice spacing.
Quantitatively, the continuum values $\Delta_2\Pi(Q^2)$ and $\Delta_2^{\mathrm{th}}\Pi(Q^2;T/2)$ differ by
only one permille.
We consider two lattices with temperature $T$ and $N_t = 20,24$ points in the Euclidean-time direction and their
equivalents with temperature $T/2$, same lattice spacing and double the points in the compact direction.
Extrapolating linearly in $a^2$ we obtain a $0.2\%$-precise estimate of the continuum value
$\Delta_2^{\mathrm{th}}\Pi(Q^2;T/2) - \Delta_2^{\mathrm{th}}\Pi(Q^2;T)$.
Considering only the finest available lattice spacing, corresponding to $N_t = 24$ for the lattice with temperature 
$T$ and $N_t^{(T/2)} = 2 N_t = 48$ for the temperature-$T/2$ one, the difference between the lattice observable
$\Delta_2^{\mathrm{th}}\mathbf{\Pi}(Q^2;T/2) - \Delta_2^{\mathrm{th}}\mathbf{\Pi}(Q^2;T)$
and its continuum value is of about $7\%$.
The lattice data, the fit curve and the continuum value of the observable are shown in the right panel 
of Figure~\ref{fig:deltaPi}.

To conclude, the free-theory analysis indicates that the strategy outlined in Section~\ref{sec:concl} to compute the 
hadronic contribution to the running of $\alpha_{\mathrm{em}}$ up to large energy scales (compared to the hadronic
scales easily accessible on the lattice) is feasible with the setup and lattice sizes suggested there.
The idea is to study the difference of hadronic vacuum polarizations $\Delta_2\Pi(Q^2)$ between the scales
$Q$ and $Q/2$ making use of thermal lattices with temperatures $T \approx Q/(8\pi)$ and $T/2$.
Having fixed $T = 246.25$~MeV and $Q = 8 \pi T \simeq 6.2$~GeV, and having considered lattices with at the most
48 points in the Euclidean-time direction, we obtained a $0.1\%$-precise estimate of the thermal quantity 
$\Delta_2^{\mathrm{th}}\Pi(Q^2;T)$.
The bias correction $\Delta_2^{\mathrm{th}}\Pi(Q^2;T/2) - \Delta_2^{\mathrm{th}}\Pi(Q^2;T)$ has been obtained with 
a precision of $0.2\%$ by extrapolating two lattice points and of $7\%$ with a single lattice spacing.

\bibliography{hvphighQ2}

\end{document}

%% file: definitions.tex
\renewcommand{\vec}[1]{\boldsymbol{#1}}
\newcommand{\be}{\begin{equation}}
\newcommand{\ee}{\end{equation}}
\newcommand{\ba}{\begin{eqnarray}}
\newcommand{\ea}{\end{eqnarray}}
\newcommand{\la}{\label}

\newcommand{\Nf}{N_\mathrm{f}}
\newcommand{\xmax}{{t}}

%% file: table_nf2.tex
\begin{table}[tb]
    \caption{Simulation parameters for the investigation with $\Nf=2$ non-perturbatively O($a$)-improved Wilson fermions.
            For the ensembles W7 and X7, the tuning to the line of constant physics defined by the temperature and the
            quark mass was performed in
            Ref.~\cite{Steinberg:2021bgr} using the Schr\"odinger functional coupling
            computed in Ref.~\cite{DellaMorte:2004bc}.
    }
    \label{tab:nf2}
    \centering
    \begin{tabular}{lc
        S[table-format=2]
        S[table-format=1.3]
        S[table-format=1.6]
        S[table-format=1.6]
        S[table-format=1.6]
        S[table-format=3]
        S[table-format=4]
        }
        \toprule
        \rule{1cm}{0pt}&\rule{1cm}{0pt}&\rule{1cm}{0pt}&\rule{2cm}{0pt}&\rule{2cm}{0pt}&\rule{2cm}{0pt}&\rule{1.5cm}{0pt}&\rule{1cm}{0pt}
        &\rule{1cm}{0pt}\\[-\arraystretch\normalbaselineskip]
        & {$T$ (MeV)} & {$L_0/a$} & {$a$ (fm)} & {$6/g_0^2$} & {$\kappa$} & {$c_\mathrm{sw}$} & {$N_\mathrm{conf}$} & {$N_\mathrm{src}$} \\
        \midrule       
        F7 & $\sim0$ & 96 & 0.0658 & 5.3      & 0.13638  & 1.90952  & 482  & 16 \\
           & 250     & 12 &        &          &          &          & 311  & 32 \\
        \midrule                                            
        O7 & $\sim0$ & 128& 0.049  & 5.5      & 0.13671  & 1.75150  & 305  & 16 \\
           & 250     & 16 &        &          &          &          & 148  & 16 \\
        \midrule                                            
        W7 & 250     & 20 & 0.039  & 5.685727 & 0.136684 & 1.64832  & 1566 & 16 \\
        \midrule                                            
        X7 & 250     & 24 & 0.033  & 5.82716  & 0.136544 & 1.58782  & 511  & 16 \\
        \bottomrule
    \end{tabular}
\end{table}

%% file: table_nf2_fits.tex
\begin{table}[tb]
  \caption{Fit ans\"atze and results for the $\Nf=2$ observables for the truncated fourth moment of the correlation function with $t=0.1974$\,fm, as
    well as for the Adler function at $Q=2.36$\,GeV.}
    \label{tab:nf2_fit}
    \centering
    \begin{tabular}{ccc
        S[table-format=1.3]
        S[table-format=1.3]
        S[table-format=1.3]
        }
        \toprule
        \rule{2cm}{0pt}&\rule{2cm}{0pt}&\rule{2cm}{0pt}&\rule{2cm}{0pt}&\rule{3cm}{0pt}&\rule{3cm}{0pt}\\[-\arraystretch\normalbaselineskip]
        observable                            & {$L_0/a$}           & ansatz  & {$c_0$ ($10^{-3}\mathrm{fm}^2$)} & {$c_1$ ($10^{-3} \mathrm{fm}$)} & {$c_2$}             \\
        \midrule       
        $\mathcal I$                          & \{96, 128\}         & L       & 1.263(7)                  & -7.0(1)               & {--}                \\
        $\mathcal I$                          & \{96, 128\}         & Q       & 1.066(4)                  & {--}                  & -0.061(1)           \\
        $\mathcal I - \mathcal I^\mathrm{th}$ & \{12, 16, 96, 128\} & L       & 0.01(1)                   & -0.4(2)               & {--}                \\
        $\mathcal I - \mathcal I^\mathrm{th}$ & \{12, 16, 96, 128\} & Q       & -0.006(6)                 & {--}                  & -0.004(1)           \\
        $\mathcal I^\mathrm{th}$              & \{12, 16\}          & Q       & 1.071(3)                  & {--}                  & -0.057(1)           \\
        $\mathcal I^\mathrm{th}$              & \{16, 20, 24\}      & Q + log & 1.053(2)                  & {--}                  & -0.148(1)           \\
        $\mathring{\mathcal I}^\mathrm{th}$   & \{16, 20, 24\}      & Q       & 1.035(2)                  & {--}                  & -0.010(1)           \\
        \midrule
                                              &                     &         & {$c_0$}                   & {$c_1$ (fm$^{-1}$)}   & {$c_2$ (fm$^{-2}$)} \\
        \midrule       
        $\mathcal D$                          & \{96, 128\}         & L       & 3.57(3)                   & -10.5(5)              & {--}                \\
        $\mathcal D$                          & \{96, 128\}         & Q       & 3.27(2)                   & {--}                  & -92(5)              \\
        $\mathcal D - \mathcal D^\mathrm{th}$ & \{12, 16, 96, 128\} & L       & 0.20(4)                   & -1.7(6)               & {--}                \\
        $\mathcal D - \mathcal D^\mathrm{th}$ & \{12, 16, 96, 128\} & Q       & 0.15(2)                   & {--}                  & -15(6)              \\
        $\mathcal D^\mathrm{th}$              & \{12, 16\}          & Q       & 3.12(1)                   & {--}                  & -76(3)              \\
        $\mathcal D^\mathrm{th}$              & \{16, 20, 24\}      & Q + log & 3.045(3)                  & {--}                  & -183(5)             \\
        $\mathring{\mathcal D}^\mathrm{th}$   & \{16, 20, 24\}      & Q       & 3.062(8)                  & {--}                  & 2(5)                \\
        \bottomrule
    \end{tabular}
\end{table}